\documentclass[%
 reprint,
superscriptaddress,
onecolumn,
 amsmath,amssymb,
 aps,
pre,
]{revtex4-2}

\usepackage{epsfig,dsfont,amssymb,amsmath,amsthm,amsfonts,amsbsy,mathrsfs}
\usepackage{graphicx}
\usepackage{color}
\usepackage{bm}
\usepackage{multirow}
\usepackage{natbib}
\usepackage{appendix}
\usepackage{changes}

\newcommand{\nn}{\nonumber }

\newcommand{\BEQ}{\begin{equation}}
\newcommand{\EEQ}{\end{equation}}
\newcommand{\BEA}{\begin{eqnarray}}
\newcommand{\EEA}{\end{eqnarray}}

\newcommand{\pt}{\partial_t}
\newcommand{\px}{\partial_x}
\newcommand{\py}{\partial_y}

\usepackage{tikz}
\usetikzlibrary{arrows,shapes,chains}

\begin{document}
\preprint{APS/123-QED}
\title{Local entropy production rate of run-and-tumble particles}

\author{Matteo Paoluzzi}

\address{Dipartimento di Fisica,
Sapienza Universit\`a di Roma, Piazzale A. Moro 2, I-00185, Rome, Italy}

\author{Andrea Puglisi}
\address{
Istituto dei Sistemi Complessi-Consiglio Nazionale delle Ricerche, and Dipartimento di Fisica,
Sapienza Universit\`a di Roma, Piazzale A. Moro 2, I-00185, Rome, Italy}

\author{Luca Angelani}
\address{
Istituto dei Sistemi Complessi-Consiglio Nazionale delle Ricerche, and Dipartimento di Fisica,
Sapienza Universit\`a di Roma, Piazzale A. Moro 2, I-00185, Rome, Italy}


\begin{abstract}
We study the local entropy production rate and the local entropy flow in active systems composed of
non-interacting run-and-tumble particles in a thermal bath.
After providing generic time-dependend expressions, we focus on the stationary regime. 
Remarkably, in this regime the two entropies are equal and depend only on the  distribution function and its spatial derivatives.
We discuss in details two case studies, relevant to real situations. 
First, we analyze the case of space-dependent speed,
describing photokinetic bacteria,  cosidering two different shapes of the speed, piecewise constant and sinusoidal.
Finally, we investigate the case of external force fields, focusing on
harmonic and linear potentials, which allow analytical treatment.
In all investigated cases, we compare exact and approximated analytical results with numerical simulations.
\end{abstract}

\maketitle
\section{Introduction}
Active matter constitutes a specific but broad class of systems within non-equilibrium physics \cite{marchetti2013hydrodynamics,bechinger2016active,elgeti2015physics}.    
Active systems are characterized by the self-propulsion of their constituent elements, whether animals, cells or artificial particles.
Such motility is powered by local injection of energy, producing non-equilibrium phenomena and the time-reversal symmetry breaking in the system. A relevant quantity for measuring the system's deviation from equilibrium is the entropy production rate (EPR) 
\cite{seifert2012stochastic,peliti2021stochastic,sarracino2025nonequilibrium}, 
which has been thoroughly studied for active particles~\cite{fodor2016far,marconi2017heat,pietzonka2017entropy,shankar2018hidden,dabelow2019irreversibility,caprini2019entropy,caprini2023entropons}.

Given the local character of the processes involved, a natural question is how entropy production can be characterized locally.
While EPR is a number associated with the entire system, it is  possible to spatially decompose EPR, making it a local quantity to gain information on specific regions where non-equilibrium fluctuations play a major role. 
This decomposition has been discussed in the past for Active Ornstein Uhlenbeck particles~\cite{marconi2017heat,puglisi2017clausius}.
It has also been introduced in the case of active field theories and, more in general, in non-equilibrium field theories \cite{PhysRevX.7.021007,PhysRevE.105.044139,PhysRevLett.133.118301,PhysRevLett.131.258302,PhysRevLett.131.258301,PhysRevE.108.064610,PhysRevE.108.064123}.
Moreover, 
EPR and local EPR have recently become experimentally measurable using model-independent techniques \cite{PhysRevLett.129.220601,di2024variance}. 

From the theoretical side, most of the attempts at computing local EPR expressions have been done within a field theoretical framework and thus working on coarse-graining descriptions where part of the information gets lost. Here we build a local EPR on the ground of the single particle dynamics. 
We focus on the run-and-tumble (RT) model \cite{Schnitzer1993,TC_2008,TC_2009,Cates_2012,martens2012probability,Angelani_2013,Detch2017},
introduced to describe the motion of motile bacteria such as {\it E.coli} \cite{Ecoli_Berg,RWinBio_Berg}, and that
has become an archetypal model for describing persistent random motions,
related to the telegrapher's equation in one space dimension \cite{Gold1951,Kac1974,Orsingher1990,Masoliver_1992,weiss2002some}.
RT models turn out to be general enough for capturing most of the active matter phenomenology, 
such as, for example, accumulation at the boundary of confining domains \cite{LiTang2009,Elgeti_2015,Elgeti2016,Angelani_2017,Malakar_2018,Ang_JPA2023,Bressloff_2023},
ratchet effect \cite{Angelani_2011,Ratchet_Reich2017,RZ2023}, 
optimal passage times \cite{Paoluzzi2020,Angelani2024EPJE,Gueneau_2024,Gueneau_2025}
and motility-induced phase separation \cite{cates2013active,MIPS2015}. 
Moreover, RT models and numerical simulations of RT particles closely reproduce the phenomenology observed in {\it E.coli} experiments \cite{Ang2009,Kurz_PRL2024,Zhao_PRE2024}.

While recent studies have addressed the computation of {\it global} EPR for RT dynamics in different situations \cite{razin2020entropy,Cocco2020,Fry2022,GMP2021,RZ2023,bao2023improving,Bressloff2024,paoluzzi2024entropy}, 
the study of {\it local} EPR remains poorly documented.
In the following, we perform a spatial decomposition of EPR considering a very general RT model that can include 
photokinetic particles, with space-dependent velocity, and the presence of external force fields, such as those generated by harmonic or linear potentials.
The expressions we obtain require knowledge of the stationary distribution as the only ingredient. 
This fact is not common as, in general, both entropy production and entropy flows require the knowledge of currents which are not entirely determined by  the distribution. Most importantly, thanks to this fact  
the analytical expressions provided here are suitable for comparison with experiments. 

Our results have a rather clear interpretation when no external force is considered and the spatial modulation affects only the velocity of the particles which is the only source of energy injection in the system: in this case we show an increase in EPR in the regions where swimmers move faster. Moreover, since active particles tend to accumulate where they move slower, the dilute areas are characterized by higher EPR. On the contrary the presence of a modulated external force produces more complicate results, as a consequence of the complex interplay between different mechanisms of energy injection. For instance, when harmonically confined, RT swimmers show local EPR distribution with a triple-peaked structure with two symmetric peaks around the accumulation points and a third peak at the trap's center (where swimmers move faster). In the presence of a (confining) piecewise linear potential, the EPR distribution appears with two symmetric peaks while there is only a central accumulation point. 

The paper is organized as follows.
In Section \ref{Sec_RT}, we introduce the run-and-tumble equations and discuss the stationary regime.
In Section \ref{Sec_EPR}, we define the local entropy rate and flow, giving general expressions of them as a function of time
and in the stationary case. 
In Section \ref{Sec_v}, we study the case of space-dependent speed, discussing piecewise constant and sinusoidal shapes.
In Section \ref{Sec_f}, we treat the case of external force fields, focusing on the harmonic and piecewise linear potentials.
Conclusions are drawn in Section \ref{Sec_Conc}.
In the Appendices, we give some details on numerical simulations (Appendix \ref{App_Sim}), 
technical calculations on solutions of steady state differential equations (Appendices \ref{App_Speed}-\ref{App_Lin})
and extend the present analysis by using different definitions of entropy (Appendix \ref{App_Entropies}).

\section{Run-and-tumble equations and stationary regime}
\label{Sec_RT}
We consider a run-and-tumble particle in one dimension in contact with a thermal bath at temperature $T$.
Let $D$ be the thermal diffusion constant, linked by the Einstein–Smoluchowski relation to the temperature, 
$D = \mu k_B T$, with $\mu$ the particle mobility, and $k_B$ the Boltzmann constant (in the following we set $k_B=1$). 
Let $v$ be the particle speed and $\alpha$ the tumbling rate at which the particle
randomly reorients its direction of motion. We consider the very general case in which the speed $v(x)$ of the particle
depends on the space variable $x$ and 
the motion occurs in regions where a generic external force field $f(x)$ is present.

While we set $\alpha$ constant in the rest of the paper, the computations in this and the next section are valid for space-varying tumbling rate $\alpha(x)$.
We indicate with $R(x,t)$ and $L(x,t)$ the probability distribution functions
(PDF), respectively of right-moving and left-moving particles, to be 
at postion $x$ at a given time $t$.
The kinetic equations governing the evolution of PDFs are \cite{paoluzzi2024entropy}
\begin{align}
    \pt R(x,t) &= - \px J_R(x,t) - \frac{\alpha}{2} (R(x,t)-L(x,t)) \ ,\\
    \pt L(x,t) &= - \px J_L(x,t) + \frac{\alpha}{2} (R(x,t)-L(x,t)) \ ,
\end{align}
where the currents are defined by
\begin{align}
\label{JR}
    J_R(x,t) &= (v(x) + \mu f(x) -D \px) R(x,t) \ ,\\
    J_L(x,t) &= (-v(x) + \mu f(x) -D \px) L(x,t) \ .
    \label{JL}
\end{align}
In terms of the quantity
\begin{align}
\label{PRL}
    P(x,t) &= R(x,t)+L(x,t) ,\\
    Q(x,t) &= R(x,t)-L(x,t) \ ,
\label{QRL}
\end{align}
the kinetic equations read
\begin{align}
    \pt P(x,t) &= - \px J(x,t) \ ,\\
    \pt Q(x,t) &= - \px \Delta(x,t)  -\alpha Q(x,t) \ ,
\label{dtQ}
\end{align}
where the global current $J$ and the difference current $\Delta$ are 
\begin{align}
     J(x,t) &= J_R(x,t)+J_L(x,t) ,\\
    \Delta(x,t) &= J_R(x,t)-J_L(x,t) \ .  
\end{align}
From (\ref{JR}) and (\ref{JL}) we can write the following relations among $P$, $Q$, $J$ and $\Delta$
\begin{align}
\label{Jdef}
    J(x,t) &= v(x)Q(x,t) +\mu f(x) P(x,t) -D \px P(x,t) \ , \\
    \Delta(x,t) &= v(x) P(x,t) +\mu f(x) Q(x,t) -D \px Q(x,t) \ .
\label{DeltaDef}
\end{align}

\subsection{Stationary regime}
In the stationary regime, i.e. for $t \to \infty$, by using (\ref{dtQ}) with $\partial_t Q=0$, (\ref{Jdef}) and (\ref{DeltaDef}),
we can write an equation for the stationary total PDF $P(x)$ (for simplicity, we use the same symbol $P$ for the time-dependent and stationary quantity, the context and explicit indication of the arguments being what determines which is appropriate) 
\begin{equation}
\label{SPDFeq0}
\px \left[ \left( v(x) - (D \px - \mu f(x)) v^{-1}(x) (D \px  -\mu f(x))  \right) P(x) \right]+ 
\frac{\alpha}{v(x)} (D\px  - \mu f(x) )P(x) = 
\left[ \px \left( D\px - \mu f(x)\right)-\alpha\right] \frac{J(x)}{v(x)} ,
 \end{equation}
or, in a different form
\begin{equation}
\label{SPDFeq0b}
\px (v(x)P(x)) - \left[ \px ( D \px - \mu f(x)) - \alpha \right]  v^{-1}(x) \left[ (D \px  -\mu f(x)) P(x)-J(x) \right] = 0 ,
\end{equation}
where the operator $\partial_x$ acts on all functions to its right.\\
In the case of vanishing stationary current, $J=0$
(we do not consider here cases in which unidirectional flows are present 
due to external forces or ratchet-like effects \cite{Angelani_2011,RZ2023}),
we have
\begin{equation}
\label{SPDFeq}
\px \left[ 
 vP(x) +\frac{\mu f(x)}{v(x)} (D \px - \mu f(x)) P(x) -D \px \left( \frac{1}{v(x)} (D \px  -\mu f(x)) P(x) \right) \right] + 
\frac{\alpha}{v(x)} (D\px  - \mu f(x) )P(x) = 0 .
\end{equation}
This is the equation for the stationary PDF $P(x)$, valid for generic space-dependent parameters $\alpha(x)$, $v(x)$ 
and force field $f(x)$. 
If we can solve (\ref{SPDFeq}), we can determine different quantities that will be useful to calculate
entropy production rates (see below). Here we give a list of such quantities as a function of $P(x)$ and its derivatives:
\begin{align}
\label{Qde}
Q(x) &= \frac{D}{v(x)} \px P(x) - \frac{\mu f(x)}{v(x)} P(x) ,\\
\Delta(x) &= v(x) P(x) + \mu f(x) Q(x) - D \px Q(x) , \label{Delde} \\
R(x) &= \frac{P(x)+Q(x)}{2} , \\
L(x) &= \frac{P(x)-Q(x)}{2} , \\
J_R(x) &= \frac{\Delta(x)}{2} , \\
J_L(x) &= -\frac{\Delta(x)}{2} .
\label{JLde}
\end{align}
After discussing, in the next section, the entropy production in the very general case,
we will treat separately the cases of space-dependet speed $v(x)$ and null force $f=0$ (section \ref{Sec_v})
and constant $v$ in the presence of a force field $f(x)$ (section \ref{Sec_f}), considering constant tumbling rate $\alpha$
in all cases.
\section{Local entropy production}
\label{Sec_EPR}
The Gibbs entropy of RT particles can be written as
\begin{equation}
    S(t) =  \int dx \ s(x,t)  \ ,
\end{equation}
where the entropy density $s(x,t)$ is the sum of the two terms for right and left-oriented particles
(recall that we set the Boltzmann constant $k_B=1$)
\begin{equation}
    s(x,t) = s_{_R}(x,t)+s_{_L}(x,t)= - R(x,t) \log R(x,t) - L(x,t) \log L(x,t) \ .
\end{equation}
We are interested in the entropy change, that for the local entropy reads
\begin{equation}
    \pt s(x,t) = - (1 + \log R(x,t) ) \pt R(x,t) - (1 + \log L(x,t)) \pt L(x,t) \ ,
\end{equation}
which, by using the kinetic equations previously derived, can be written as follows
\begin{equation}
\label{ler}
    \pt s(x,t) = \pi(x,t) -\phi(x,t) \ .
\end{equation}
The first term is the local entropy production rate, 
which is a non-negative quantity and turns out to be
\begin{equation}
    \pi(x,t) = \frac{1}{D} \left( \frac{J_R^2(x,t)}{R(x,t)} + \frac{J_L^2(x,t)}{L(x,t)}\right) + 
\frac{\alpha}{2} (R(x,t)-L(x,t)) \log \frac{R(x,t)}{L(x,t)} \ .
\label{pid}
\end{equation}
It is worth noting that the above expression can be written as the sum of three contributions
\begin{equation}
\label{pisum}
    \pi(x,t) = \pi_R(x,t) + \pi_L(x,t) + \pi_{RL}(x,t) \ ,
\end{equation}
where the first two terms are contributions of the right-oriented and left-oriented particles and the last one is the 
relative entropy (or Kullback–Leibler divergence)
\begin{align}
    \pi_R(x,t) &= \frac{1}{D} \frac{J_R^2(x,t)}{R(x,t)} , \\
    \pi_L(x,t) &= \frac{1}{D} \frac{J_L^2(x,t)}{L(x,t)} , \\
    \pi_{RL}(x,t) &= \frac{\alpha}{2} (R(x,t)-L(x,t)) \log \frac{R(x,t)}{L(x,t)} .
\end{align}
The second term in (\ref{ler}) is the local entropy flow 
(to the environment and surrounding regions)
\begin{equation}
\label{phid}
    \phi(x,t) = \frac{v(x)}{D} (J_R(x,t) - J_L(x,t)) +\frac{\mu f(x)}{D} (J_R(x,t) + J_L(x,t)) 
    - \px \chi(x,t) \ ,
\end{equation}
where $\chi$ is defined by
\begin{align}
\label{chid}
    \chi(x,t) &= J_R(x,t) (1+\log R(x,t)) + J_L(x,t) (1+\log L(x,t)) \nonumber \\ 
    &= 
    (J_R(x,t)+J_L(x,t)) \left( 1 + \frac12 \log R(x,t)L(x,t) \right) +
    \frac{J_R(x,t)-J_L(x,t)}{2} \log \frac{R(x,t)}{L(x,t)}\ .
\end{align}
From these local entropy terms we can obtain the global entropies by integrating over the space variable $x$.
The global entropy production rate $\Pi$ turns out to be
\begin{equation}
\label{Pi}
    \Pi(t) = \int dx\ \pi(x,t) =\int dx \left[  \frac{1}{D} \left( \frac{J_R^2(x,t)}{R(x,t)} + 
    \frac{J_L^2(x,t)}{L(x,t)}\right) + 
\frac{\alpha}{2} (R(x,t)-L(x,t)) \log \frac{R(x,t)}{L(x,t)} \right]\ ,  
\end{equation}
and the global entropy flux $\Phi$
\begin{equation}
\label{Phi}
    \Phi(t) =\int dx\ \phi(x,t) =\int dx \left[ \frac{v(x)}{D} (J_R(x,t) - J_L(x,t)) +
    \frac{\mu f(x)}{D} (J_R(x,t) + J_L(x,t))  \right]\ ,
\end{equation}
having assumed vanishing distributions at boundaries.
We note that the last derivative term in (\ref{phid}) does not contribute to the global entropy flux, 
but must still be considered when calculating the local entropy.\\
It is interesting to note that (\ref{ler}) can be written as a local balance equation for the entropy
\begin{equation}
    \pt s(x,t) + \px J_s(x,t) = \pi(x,t) + \varphi(x,t) , 
\end{equation}
where the quantity $J_s(x,t)=-\chi(x,t)$ can be interpreted as the entropy current and 
$\varphi(x,t)=-\phi(x,t)-\px \chi(x,t)$
is the  entropy flow from the environment, given by (minus) the first two terms in the RHS of (\ref{phid}).

\subsection{Thermodynamic interpretation}

A thermodynamic interpretation of $\varphi(x,t)$ follows from a straightforward manipulation of Eq.~\eqref{phid}:
\begin{align}
\label{phi2}
\varphi(x,t) &= - \frac{1}{\mu T}\left[ (v(x)+\mu f(x))J_R(x,t) + (-v(x)+\mu f(x))J_L(x,t)\right] \nonumber \\
&=\frac{1}{T}(\dot{\mathcal{Q}}_R(x,t)+\dot{\mathcal{Q}}_L(x,t)) ,
\end{align}
where we have introduced the local densities of heat exchange rates for the right (left) populations $\dot{\mathcal{Q}}_{R(L)}$, and (implicitly) the right (left) thermal bath forces $F^B_{R(L)}$ :
\begin{subequations} \label{therm}
\begin{align} 
    \dot{\mathcal{Q}}_R(x,t) &= -\frac{1}{\mu}(v(x)+\mu f(x)) J_R(x,t)=F^B_R(x,t)J_R(x,t)\\
    \dot{\mathcal{Q}}_L(x,t) &= -\frac{1}{\mu}(-v(x)+\mu f(x)) J_L(x,t)=F^B_L(x,t) J_L(x,t)\\
    \mu F^B_R(x,t)&=-(v(x)+\mu f(x)) =-\dot{x}+\eta(t) \\
    \mu F^B_L(x,t)&=-(-v(x)+\mu f(x)) =-\dot{x}+\eta(t), 
\end{align}
\end{subequations}
see Eq.~\eqref{Lang_eq} in the Appendix \ref{App_Sim} for an explanation of the (intuitive) meaning of $\dot{x}$ and $\eta(t)$. Summarizing, Eq.~\eqref{phi2} suggests to interpret $\varphi(x,t)$ as the usual Clausius term for the entropy change caused by heat exchanges with the external bath.
Integrating along all space leads to $\int dx\  \varphi(x,t) = \dot{\mathcal{Q}}(t)/T$ having defined $\dot{\mathcal{Q}}(t)=\int dx (\dot{\mathcal{Q}}_R(x,t)+\dot{\mathcal{Q}}_L(x,t))$.\\

\subsection{Stationary regime}
In the stationary regime we have that $\pt s=0$ and, then, the two local entropy rates
(\ref{pid}) and (\ref{phid}), which are now only space-dependent quantities, 
turn out to be equal
\begin{equation}
\label{equiv}
    \pi (x) = \phi(x) ,
\end{equation}
as well as the corresponding integrated quantities (\ref{Pi}) and (\ref{Phi})
\begin{equation}
    \Pi = \Phi .
\end{equation}
By assuming vanishing stationary current $J=0$, we can write the local EPR (\ref{pid}) as
\begin{align}
\label{pi_1}
  \pi(x) &=   \frac{1}{D} \left( \frac{J_R^2(x)}{R(x)}+ \frac{J_L^2(x)}{L(x)} \right)+ 
  \frac{\alpha}{2} (R(x)-L(x))  \log \frac{R(x)}{L(x)} \nonumber \\
  &= \frac{\Delta^2(x)}{4D} \left( \frac{1}{R(x)} +\frac{1}{L(x)} \right) + \frac{\alpha Q(x)}{2}  \log \frac{R(x)}{L(x)}\nonumber  \\ 
  &= \frac{\Delta^2(x)}{D} \frac{P(x)}{P^2(x)-Q^2(x)} + \frac{\alpha Q(x)}{2} \log \frac{P(x)+Q(x)}{P(x)-Q(x)} .
\end{align}
This can be expressed as a function of the stationary PDF $P(x)$ and its derivaties, 
by exploiting (\ref{Qde}-\ref{JLde}).\\
Alternatively, using (\ref{phid}) and (\ref{equiv}), we can write the local EPR as $\pi(x)=\phi(x)$, where
\begin{align}
\label{pi_2}
    \phi(x) & = \frac{v(x)}{D} (J_R(x) - J_L(x)) - \px \left( \frac{J_R(x)-J_L(x)}{2} \log \frac{R(x)}{L(x)}\right) \nonumber \\
    &= \frac{v(x) \Delta(x)}{D}- \frac12 \px \left( \Delta(x) \log \frac{R(x)}{L(x)}\right)  \nonumber \\
    &= \frac{v(x) \Delta(x)}{D}- \frac12 \px \left( \Delta(x) \log \frac{P(x)+Q(x)}{P(x)-Q(x)}\right) .
\end{align}
Again, we can express this quantity as a function of the stationary PDF $P(x)$ 
and its derivatives through (\ref{Qde}-\ref{JLde}).
Indeed, we obtain
\begin{align}
\label{LEPR_P}
    \pi(x) = \sigma(x) - \partial_x \chi(x) ,
\end{align}
where 
\begin{align}
\label{LEPR_delta}
    \sigma(x) &= \frac{v^2(x)-\mu^2f^2(x)}{D} P(x) + 2\mu f(x) \px P(x) + \mu P(x) \px f(x) -D \px^2 P(x) \nonumber\\
    &+ 
    D (\px P(x))(\px \log v(x)) -\mu f(x) P(x) \px \log v(x) ,\\
    \chi(x) &= \frac{D\sigma(x)}{2v(x)} \ \log  \frac{v(x)P(x)-\mu f(x)P(x) +D\px P(x)}{v(x)P(x)+\mu f(x)P(x) -D\px P(x)} .
\label{LEPR_chi}
\end{align}
We emphasize that, as in the previous section, the expressions obtained are quite general,
valid for generic space-dependent tumbling rate $\alpha(x)$, speed $v(x)$ and force field $f(x)$.
In the following sections we will analyze the non-homogeneous speed and force field cases separately. 
We observe that, in general, the computation of $P(x)$ is quite easy (or at least not too difficult in principle) 
to perform in numerical simulations and experiments, but it is challenging to find analytical expression for it.
However, in the following we will show that we are able to provide explicit expressions for $P(x)$, and therefore for $\pi(x)$,
in some special cases, such as piecewise constant velocity profiles,  harmonic and linear potentials.

\section{Velocity field}
\label{Sec_v}
We consider here the case of a free ($f=0$) RT particle with constant tumbling rate $\alpha$ and a space-dependent speed $v(x)$ \cite{AngGar2019,Sandev_2025}.
The typical application we have in mind is the case of photokinetic bacteria whose velocity can be controlled using arbitrary light patterns \cite{Elife2018,pellicciotta2023colloidal}. Since $v(x)$ is tunable as an external parameter and $P(x)$ can be computed from experimental data, it is an ideal set-up for measuring $\pi(x)$ without estimating the Kulback-Leibler divergence between a trajectory and its time reversal.\\
The equation for the stationary PDF reads
\begin{equation}
\label{SPDFeq_v}
\px \left[ 
v(x)P(x)  -D^2 \px \left( \frac{\px P(x)}{v(x)}\right) \right]
+ \frac{\alpha D}{v(x)} \px P(x) =0 \ ,
\end{equation}
as can be see by putting $f=0$ in the equation (\ref{SPDFeq}) valid for null net stationary current $J=0$.
For purely active particles, i.e., in the absence of thermal noise $D=0$, we have that $\partial_x (vP)=0$ and 
then we obtain the well known $1/v$ dependence of the stationary PDF
$$P(x) \propto  \frac{1}{v(x)} , \qquad (D=0).$$
The local EPR can be expressed in terms of $P$ and its derivatives.
By using (\ref{pi_1}) we have
\begin{align}
\label{pi_v0}
\pi(x) &= \frac{P(x)}{D} \frac{\Delta^2(x)}{P^2(x)-Q^2(x)} + \frac{\alpha}{2}Q(x) \log \frac{P(x)+Q(x)}{P(x)-Q(x)} , \\ 
    Q(x) &= D\frac{\px P(x)}{v(x)}     \ , \\
    \Delta(x) & = v(x)P(x) - D^2 \px\left( \frac{\px P(x)}{v(x)}\right) \ .
\end{align}
Alternatively, thorugh the relation (\ref{LEPR_P}), $\pi=\phi$,  and  using  (\ref{LEPR_delta}) and (\ref{LEPR_chi}) 
we can write
\begin{align}
\label{pi_v}
\pi(x) &=  \sigma(x) - \partial_x \chi(x) , \\ 
    \sigma(x) &= \frac{v^2(x)}{D} P(x) - D \px^2 P(x) + D (\px P(x)) (\px \log v(x))    \ , \\
    \chi(x) & = \frac{D\sigma(x)}{2v(x)} \ \log \frac{v(x)P(x)+D\partial_x P(x)}{v(x) P(x)-D\partial_x P(x)} \ .
\end{align}

\paragraph*{Perturbative expansion.}
\label{Pertex}
Exact solutions of (\ref{SPDFeq_v}) are not always possible, and it is often useful to follow a perturbative approach 
for small values of diffusivity.
In many real cases and practical realization of active systems 
(e.g, motile cells or syntetic catalytic particles) 
the thermal diffusivity is some order of magnitude lower than the active one.
We introduce the dimensionless quantity 
\begin{equation}
    D_* = \frac{\alpha D}{v_0^2} \ ,
\end{equation}
which is the ratio $D/D_A$ between thermal ($D$) and active diffusion ($D_A=v_0^2/\alpha$),
where $v_0$ sets the scale of velocity,  $v(x) = v_0 w(x)$, with
$w(x)$ a dimensionless field. The quantity $D_*$ is the inverse of the Peclet number
$Pe=v_0^2/(\alpha D)$.
We look for the expressions of PDFs and EPRs in the limit
$D_* \ll 1$ (large Peclet number $Pe\gg 1$).
The differential equation (\ref{SPDFeq_v}) for $P$ can be rewritten in the form 
\begin{equation}
\label{FP_Pst}
    \partial_x \left[
w(x) P(x) - \frac{v_0^2}{\alpha^2} D_*^{2} \partial_x \left( \frac{\partial_x P(x)}{w(x)} \right)
    \right]
    + \frac{D_*}{w(x)} \partial_x P(x) = 0 \ .
\end{equation}
By assuming that  $P$ is an analytic function of $D_*$ we can write the series expansion
\begin{equation}
\label{Pser}
    P(x) = \sum_{n=0}^\infty P_n(x) \ D_*^{n} \ ,
\end{equation}
where the $P_n(x)$ are functions that do not depend on $D_*$ and can take negative values (except $P_0$ which is non-negative defined).
The normalization condition implies that
\begin{equation}
\label{norm}
    \int dx \ P_n(x) = \delta_{n0} ,
\end{equation}
where $\delta_{nk}$ is the Kronecker delta function ($1$ for $n=k$ and $0$ otherwise).
The differential equations for $P_n$ are obtained by considering (\ref{FP_Pst}) 
at different orders of $D_*$
\begin{align}
     &\partial_x(w(x) P_0(x)) = 0 \ ,   &(n=0) \\
     &\partial_x(w(x) P_1(x)) + \frac{1}{w(x)} \partial_x P_0(x) = 0 \ ,   & (n=1) \\
     &\partial_x \left[ w(x) P_n(x) - \frac{v_0^2}{\alpha^2} 
    \partial_x \left( \frac{\partial_x P_{n-2}(x)}{w(x)}\right)   \right]
    + \frac{1}{w(x)} \partial_x P_{n-1}(x) = 0  ,    &(n\geq 2) 
\end{align}
By solving up to the second order we have
\begin{eqnarray}
\label{Pper_0}
P_0(x) &=& \frac{c_0}{w(x)} \ , \\
\label{Pper_1}
P_1(x) &=& \frac{c_1}{w(x)} - \frac{c_0}{2 w^3(x)} \ , \\
P_2(x) &=& \frac{c_2}{w(x)} - \frac{c_1}{2 w^3(x)} + \frac{3 c_0}{8 w^5(x)} 
- \frac{c_0 v_0^2}{\alpha^2 w(x)} 
\partial_x \left( \frac{\partial_x w(x)}{w^3(x)}\right) \ .
\label{Pper_2}
\end{eqnarray}
The constants $c_n$ are determined by the normalization conditions (\ref{norm})
(we assume that the motion occurs in a finite domain of size $L$ with periodic boundary conditions)
\begin{eqnarray}
c_0 &=& \frac{1}{L} \langle w^{-1}(x) \rangle^{-1} \ , \\
c_1 &=& \frac{1}{2L} \frac{\langle w^{-3}(x) \rangle}{\langle w^{-1}(x) \rangle^2}\ ,  \\
c_2 &=& \frac{1}{L} \frac{1}{\langle w^{-1}(x) \rangle^2} \left[ 
\frac14 \frac{\langle w^{-3}(x) \rangle^2}{\langle w^{-1}(x) \rangle} -
\frac38 \langle w^{-5}(x) \rangle +
\frac{v_0^2}{\alpha^2} \langle (\partial_x w(x))^2 w^{-5}(x) \rangle 
\right] \ ,
\label{cn}
\end{eqnarray}
where 
\begin{equation}
    \langle f(x) \rangle = \frac{1}{L} \int dx \ f(x)\ .
\end{equation}

We now analyze the entropy production in the small diffusion regime.
We can express the EPR as a Laurent series in $D_*$
\begin{equation}
    \pi(x) = \sum_{n=-1}^\infty \pi_n(x) \ D_*^n ,
\end{equation}
where the term $n=-1$ represents the singular part.
By using the expansion of $P$ (\ref{Pser}) inserted in (\ref{pi_v}) we obtain, 
up to the first order in $D_*$
\begin{equation}
\label{EPRd}
    \pi(x) =\frac{\pi_{-1}(x)}{D_*} + \pi_0(x) + \pi_1(x)\ D_* + O(D_*^2) \ ,
\end{equation}
where
\begin{eqnarray}
\label{EPRd_0}
    \pi_{-1}(x) &=& \alpha c_0  w(x) \ ,  \\
 \label{EPRd_1}
   \pi_{0}(x) &=& \alpha \left[ c_1 w(x) - \frac{c_0}{2 w(x)} \right]\ , \\
 \label{EPRd_2}
   \pi_{1}(x) &=& \alpha \left[ c_2 w(x) - \frac{c_1}{2 w(x)} +\frac{3c_0}{8 w^3(x)} \right] 
    + \frac{c_0v_0^2}{\alpha} \partial_x \left( \frac{\partial_x w(x)}{w^2(x)}\right)  \ ,
\end{eqnarray}
with the constants $c_n$ given in (\ref{cn}). 
In the limit of small $D$ one has
\begin{equation} \label{smallD}
    D \pi(x) \approx \frac{D \pi_{-1}(x)}{D_*}=v_0^2 c_0 w(x)=\frac{v(x)}{L \langle v(x)^{-1}\rangle} ,
\end{equation}
and, noting that in the same limit $P(x)\approx P_0(x)=c_0/w(x)$, we have
$D\pi(x) \approx v^2(x) P(x)$ and then
\begin{equation}
\label{smallD2}
D \pi(x) \approx \frac{1}{P(x)L^{2}\langle v(x)^{-1}\rangle^{2}} .
\end{equation} 
The above expressions show that, for small $D$, the EPR profile has a direct monotonicity relation with the velocity field and an inverse one with the density profile.

The total EPR is obtained by integrating the EPR density (\ref{EPRd}) over space, giving 
\begin{equation}
   \Pi = \frac{\Pi_{-1}}{D_*} + \Pi_0 + \Pi_1 \ D_* + O(D_*^{2}) \ ,    
\end{equation}
with
\begin{eqnarray}
    \Pi_{-1} &=& \alpha \frac{\langle w(x) \rangle}{\langle w^{-1}(x) \rangle} \ ,  \\
    \Pi_{0} &=& \frac{\alpha}{2} \left[ 
    \frac{\langle w(x) \rangle \langle w^{-3}(x) \rangle }{\langle w^{-1}(x) \rangle^2} - 1
    \right] \ , \\
    \Pi_1 &=& \alpha L c_2 \langle w(x) \rangle + \frac{\alpha}{8} 
    \frac{\langle w^{-3}(x) \rangle}{\langle w^{-1}(x)\rangle} \ ,
\end{eqnarray}
where the constant $c_2$ in the last expression is given in (\ref{cn}).

\subsection{Piecewise constant velocity}
We analyze here the case of a gas of RT particles immersed in a piecewise constant speed field
\begin{equation}
v(x)= \begin{cases}
v_1  & \qquad \text{if $0<x<\lambda$}\\
v_2 & \qquad \text{if $\lambda<x<L$}
\end{cases}
\end{equation}
moving in a box of size $L$ with periodic boundary conditions. 
In this case the problem admits of an exact solution.
To find the stationary PDF we have to solve equation (\ref{SPDFeq_v}) in the two zones
$0<x<\lambda$ ($1$) and $\lambda<x<L$ ($2$), imposing the appropriate boundary conditions.
Within the two zones the speed is constant, $v(x)=v_i$ ($i=1,2$), and the equations for the PDF read
\begin{equation}
\px \left[ \left(v_i^2-D^2\px^2 \right) P_i(x)\right] + \alpha D \px P_i(x) = 0 , \qquad (i=1,2) ,
\end{equation}
whose solutions are
\begin{equation}
    P_i(x) = A_i e^{k_i x} + B_i e^{-k_i x} + C_i ,
\end{equation}
with
\begin{equation}
\label{k_i}
    D^2 k_i^2 = v_i^2 +\alpha D .
\end{equation}
The six unknown constants $\{A_i,B_i,C_i\}\  (i=1,2)$
are determined by imposing suitable conditions on the solution (see Appendix \ref{App_Speed}).\\
Once we know the expression of $P(x)$ we can calculate the local EPR through  
(\ref{pi_v0}) or (\ref{pi_v}) considered, separately, in the two zones of constant speed.
By noting that the quantities $P$, $Q$ and $\Delta$ are continuous at the points where 
the speed jumps, $x=0$ and $x=\lambda$ (see Appendix \ref{App_Speed}), also $\pi$ is. Fig. (\ref{fig:rt_step}) reports the comparison between numerical simulations (see Appendix \ref{App_Sim}) and analytical solutions for both $P(x)$ and $\pi(x)$. 
For all values of $D$ one observes that $\pi(x)$ decreases where particles are slower i.e. where $v(x)$ decreases or, equivalently, where $P(x)$ increases. For small values of $D$ one can also check that the leading order approximation, Eq.~\eqref{smallD}, is well reproduced, as $\langle v^{-1}(x)\rangle=3/2$ and therefore $D\pi(x)$ goes from $1/3$ to $2/3$. 


\begin{figure}[!t]
\centering\includegraphics[width=1.\textwidth]{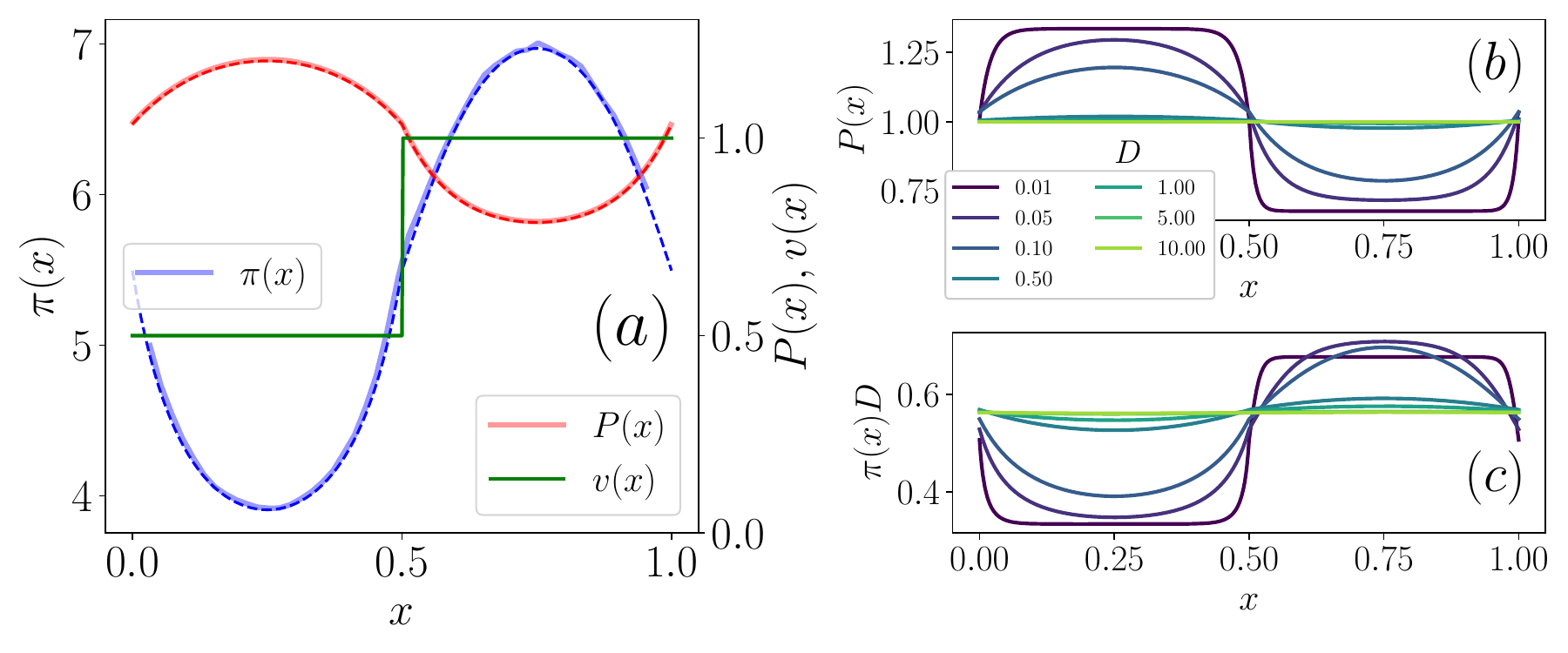}
\caption{Run-and-tumble particles with piecewise constant velocity field. (a) Local entropy production rate (blue) and stationary distribution (red) of RT immersed in a piecewise constant speed with $v_1=1/2$ (for $x<\lambda=1/2$) and $v_2=1$ (for $1/2<x<L=1$). Dashed lines refer to theoretical predictions, solid lines are numerical simulations ($\alpha=1$ and $D=0.1$). (b) Analytical expression of $P(x)$ as $D$ increases (increasing values from violet to yellow, see legend) (c) The corresponding local entropy production rate $\pi(x) D$.}
\label{fig:rt_step}      
\end{figure}

\subsection{Sinusoidal velocity}
\begin{figure*}[!t]
\centering\includegraphics[width=1.\textwidth]{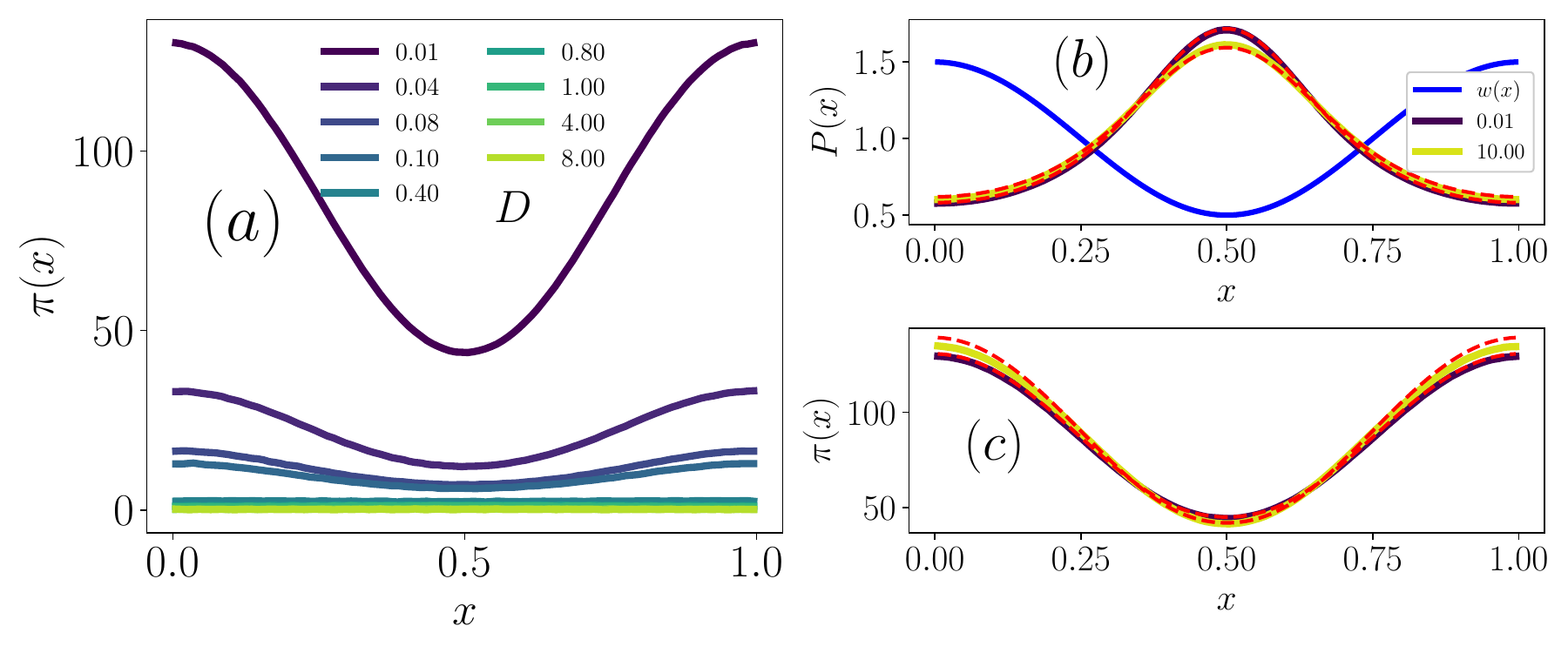}
\caption{Run-and-tumble particles with sinusoidal velocity field. (a) Local entropy production rate $\pi(x)$ from numerical simulations ($\alpha=1$, $v_0=1$, $L=1$)
for increasing values of $D$ (from violet to yellow, see legend). (b) Stationary distribution computed from numerical simulations (different colors indicate different $\alpha$ values, see the legend, for $D=10^{-2},v=1$), the dashed red lines are $P(x)$ computed using perturbation theory ($\alpha=0.01,10$). The solid blue line is the velocity profile.
(c) Local entropy production rate $\pi(x)$ from numerical simulations (same color code as in (b)) and using perturbation theory (dashed red lines as in (b)).}
\label{fig:rt_cos}      
\end{figure*}

We now analyze the case of a periodic sinusoidal shape of the particle speed 
\begin{equation}
    v(x) = v_0 \left(1 +\frac12 \cos \frac{2\pi x}{L} \right) ,
\end{equation}
where $L$ is the spatial period. The minimum and maximum speed values are $v_{min}=v_0/2$ and $v_{max}=3v_0/2$, see the profile of $w(x)=v(x)/v_0$ in Fig. (\ref{fig:rt_cos})b.
In this case it is not easy to obtain exact expressions of the PDF $P(x)$ for generic $D$ 
and we have to resort to perturbative analysis previously developed, valid for small $D$.
Perturbative expressions of $P(x)$ can be obtained as a function of $D_*=\alpha D/v_0^2$ up to the second order by using
(\ref{Pser}) and (\ref{Pper_0})-(\ref{Pper_2}). For the local EPR $\pi(x)$ we can use (\ref{EPRd}) and (\ref{EPRd_0})-(\ref{EPRd_2}).

Fig. (\ref{fig:rt_cos}) reports a comparison between numerical simulations and perturbative computation. 
Panel (a) shows the behavior of $\pi(x)$ for increasing values of $D$ at fixed $\alpha=1$. Panel (b) and (c) show what happens when $\alpha$ is changed from $0.01$ to $10$ at fixed small $D=10^{-2}$, in particular panel (b) displays the  density profiles $P(x)$, while panel (c) shows $\pi(x)$. As in the case of piecewise constant velocity, we observe a kind of ``inverse monotonicity'' between $P(x)$ and $\pi(x)$ -- see Eq. (\ref{smallD2}) -- which is qualitatively explained by the expectation that $\pi(x)$ is larger where $v(x)$ is larger and therefore where  $P(x)$ is smaller. Again, in the limit of small $D$, we can verify the fair prediction of Eq.~\eqref{smallD}, considering that $\langle v(x)^{-1} \rangle= 2/\sqrt{3} \approx 1.155$ and therefore $\pi(x)$ oscillates between $\sim 43$ and $\sim 130$.

We stress that these results refer to the case of non-interacting bacteria.
A natural question is what happens once we consider forces as obstacles since, in this case, feedback between local density and local velocity can lead to phase separation.

\section{Force field}
\label{Sec_f}
To understand the effect of forces on local EPR we focus on simple situations with external fields so that we still are dealing with a gas of RT particles.
We consider here the case of a RT particle with constant tumbling rate $\alpha$ and speed $v$ 
in the presence of a force field $f(x)$.
The equation for the steady state PDF takes the form -- see (\ref{SPDFeq}) with constant $v$ 
\begin{equation}
\label{eq_st}
    \px \left[ \left( v^2 - (D\px - \mu f(x))^2 \right) P(x) \right] +
    \alpha (D\px -\mu f(x)) P(x) =0 \ .
\end{equation}
It is interesting to analyze the two opposite limits of purely Brownian particles and purely
active particles in the stationary regime.
\\
\paragraph*{Brownian limit.}
In the absence of active motion ($v=0$) the PDF $P_B$ of the Brownian particle obeys the equation
\begin{equation}
\label{eq_PB}
    (D \px -\mu f(x)) P_B(x) = 0 \ ,
\end{equation}
as simply obtained from (\ref{Jdef})   considering null flux $J=0$.
By introducing the potential $V$, $f(x)=-\px V(x)$, 
the solution is 
\begin{equation}
    P_B(x) = C_B \ e^{-\frac{\mu}{D} V(x)} \ ,
\end{equation}
where $C_B$ is the normalization constant
\begin{equation}
    C_B = \left[ \int dx \ e^{-\frac{\mu}{D}V(x)}\right]^{-1} \ .
\end{equation}
\\
\paragraph*{Active limit.}
In the opposite limit of purely active motion ($D=0$) the stationary equation 
for the PDF $P_A$ reads
\begin{equation}
\label{eq_PA}
  \px \left[ (v^2 -\mu^2 f^2(x)) P_A(x) \right] - \alpha \mu f(x) P_A(x) = 0 \ ,
\end{equation}
whose solution can be written for $v>\mu f(x)$
\begin{equation}
    P_A(x) = \frac{C_A}{v^2-\mu^2 f^2(x)} \ 
    \exp \left[\alpha \mu  \int^x dy \ \frac{f(y)}{v^2-\mu^2 f^2(y)} \right]\ ,
\end{equation}
where $C_A$ is a normalization constant \cite{Dhar2019,Sevilla_2019}.\\

The local EPR can be expressed in terms of $P$ and its derivatives -- see (\ref{LEPR_P}) with constant $v$
\begin{equation}
\label{EPRl}
    \pi(x) = \frac{1}{D} \left[ v^2 - \left( D \px -\mu f(x)\right)^2\right] P(x)    
    - \px \chi(x) \ ,
\end{equation}
where 
\begin{equation}
    \chi(x) = \frac{1}{2v} 
    \left[ \log \frac{(v-\mu f(x) +D\px)P(x)}{(v+\mu f(x)-D\px)P(x)} \right] 
    \left[v^2 - \left( D\px-\mu f(x)\right)^2 \right] P(x) \ .
\end{equation}
It is noteworthy to point out that we were able to express the local EPR $\pi(x)$
as a function of the stationary PDF $P(x)$, Eq. (\ref{EPRl}), allowing us to obtain 
explicit expressions of it whenever we know the stationary solution of the problem.  
However, it should be added that it is not always possible to obtain explicit expressions 
of the stationary distributions, except in a limited number of cases, such as the case of confined motions in a box~\cite{razin2020entropy} and the case of harmonic and piecewise linear potentials, 
that we discuss in the following Sections.

\subsection{Harmonic potential}
\label{HarPot}
The case of harmonic potential is of particular interest because of its physical relevance and 
analytical tractability.
The potential $V(x)$ and the force $f(x)=-\px V(x)$ are given by
\begin{align}
V(x) &= \frac{k}{2} x^2  \ , \\
f(x) &= - k x \ .
\end{align}
\\
\paragraph*{Stationary solution.}
In this case, it is possible to write the solution of (\ref{eq_st}) as a convolution of
Brownian and active PDFs
\begin{equation}
\label{P_harm}
    P(x) = \int_{-\ell}^{+\ell} dy \ P_A(y) \ P_B(x-y) \ ,
\end{equation}
where $\ell = v/\mu k$ \cite{Frydel2022}.
In the above expression, the Brownian PDF is
\begin{equation}
\label{P_B}
    P_B(x) = \sqrt{\frac{\mu k}{2\pi D}} \ e^{-\frac{\mu k}{2 D}x^2}
\end{equation}
and the active one is given by
\begin{equation}
\label{P_A}
    P_A(x) = C \left(v^2 - \mu^2 k^2 x^2 \right)^{\frac{\alpha}{2\mu k}-1} \ ,
\end{equation}
where the constant $C$ is
\begin{equation}
\label{P_AA}
    C = \frac{\mu k \Gamma\left(\frac{\alpha}{2\mu k}+\frac12\right)}{\sqrt{\pi}\Gamma\left(\frac{\alpha}{2\mu k}\right)} \ v^{1-\frac{\alpha}{\mu k}}
\end{equation}
with $\Gamma(z)$ the Euler gamma function.
The proof of (\ref{P_harm}) is given in the Appendix \ref{App_Harm}.\\
Some limiting cases are listed below.\\

\paragraph{Diffusive Limit of the Active Motion.} For $\alpha,v \to \infty$ with finite $D_A=v^2/\alpha$ (active diffusivity)
the active PDF (\ref{P_A}) tends to a Gaussian
\begin{equation}
    P_A(x) = \sqrt{\frac{\mu k }{2 \pi D_A}}\ e^{-\frac{\mu k}{2D_A} x^2} ,
\end{equation}
and then (\ref{P_harm}) is a convolution of two Gaussian distributions ($\ell = v/ \mu k \to \infty$) 
resulting in a Gaussian with effective variance $\sigma^2$
\begin{equation}
    P(x) =  \frac{1}{\sqrt{2 \pi \sigma^2}}\ e^{-\frac{x^2}{2\sigma^2}} ,
\end{equation}
with
$$
\sigma^2 = \frac{D +D_A }{\mu k} . 
$$
\paragraph{Ballistic Limit.} For $\alpha\to 0$ the active PDF has two delta peaks
\begin{equation}
    P_A(x) = \frac12 \left[ \delta \Big(x-\frac{v}{\mu k} \Big) + \delta \Big(x+\frac{v}{\mu k}\Big)  \right]
\end{equation}
and the total PDF is the sum of two Gaussian distributions
\begin{equation}
    P(x) = \frac12 \sqrt{\frac{\mu k}{2\pi D}} \left(     
    e^{-\frac{\mu k }{2D}(x-v/\mu k)^2} +
    e^{-\frac{\mu k }{2D}(x+v/\mu k)^2} 
    \right).
\end{equation}

\begin{figure*}[!t]
\centering\includegraphics[width=1.\textwidth]{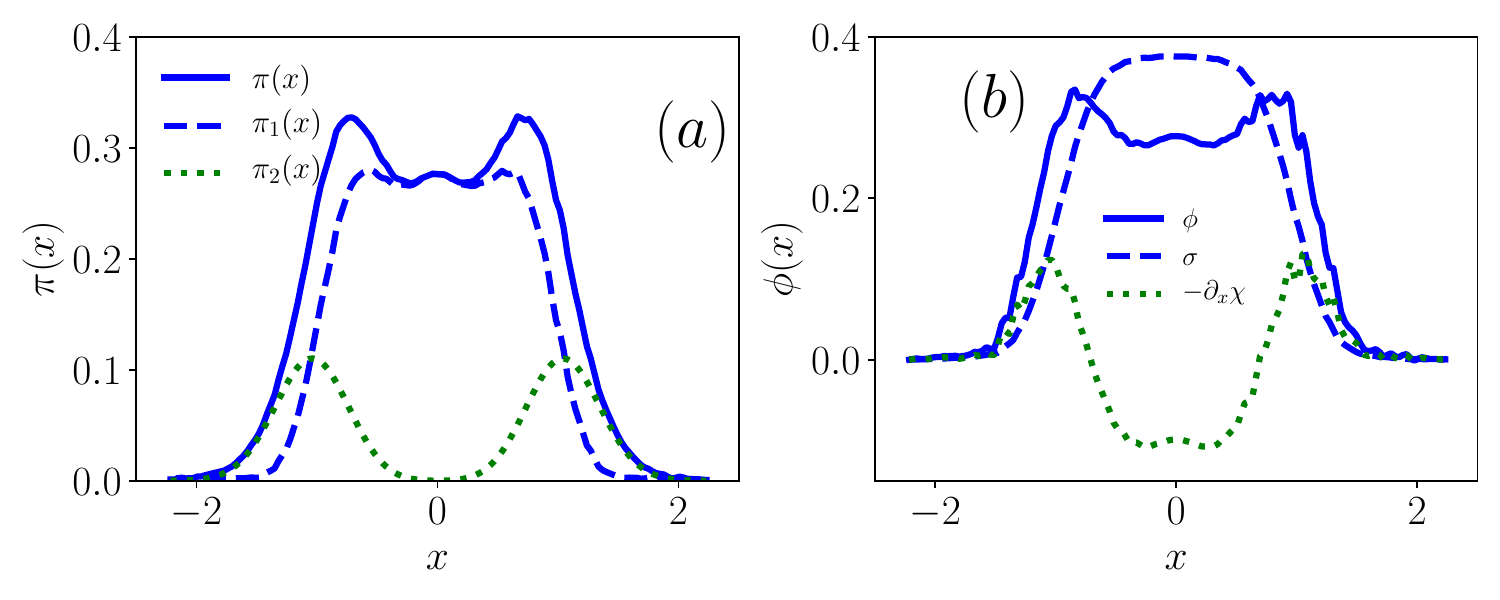}
\caption{Run-and-tumble particle in a harmonic trap. (a) Local entropy production rate $\pi(x)$ computed through (\ref{pi_1}) from numerical simulations, with $\pi_1(x) \equiv \frac{\Delta^2}{4 D}(\frac{1}{R} + \frac{1}{L})$ and $\pi_2(x) \equiv \frac{\alpha Q}{2} \log \frac{R}{L}$. (b) $\phi(x)=\pi(x)$ computed from numerical data through (\ref{pi_2}) ($\alpha=0.08$, $\mu=v=k=1$, and $D=0.1$).}
\label{fig:rt_harm1}      
\end{figure*}

\begin{figure*}[!t]
\centering\includegraphics[width=1.\textwidth]{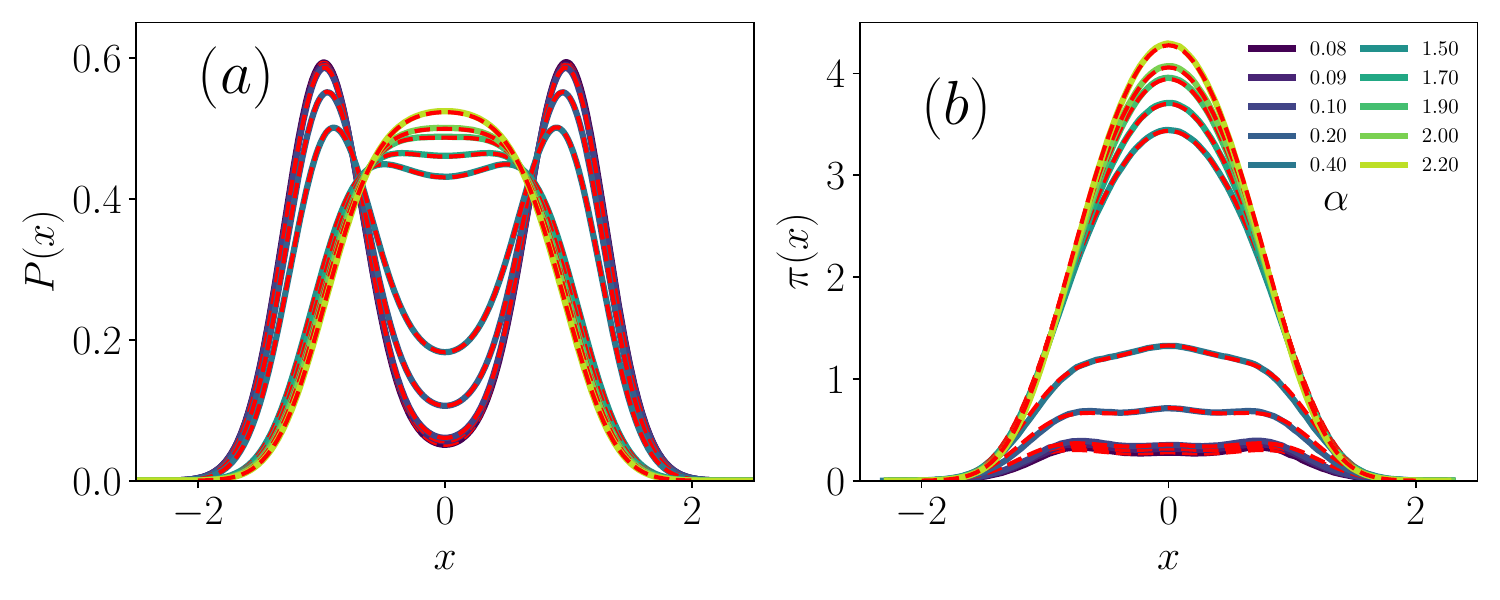}
\caption{Run-and-tumble particle in a harmonic trap. (a) Stationary distribution: comparison between numerical simulations (solid lines) and (\ref{P_harm}) ($\alpha \in [0.08,2.20]$ (see legend in (b)), $\mu=v=k=1$, and $D=0.1$). (b) $\pi(x)$ computed in numerical simulations (solid line) compared with (\ref{eprharmexpl}). Different colors refer to different $\alpha$ values (see legend). We consider $\mu=v=k=1$, and $D=0.1$.}
\label{fig:rt_harm2}      
\end{figure*}

\paragraph*{Entropy production rate.}
In the harmonic case the local EPR (\ref{EPRl}) becomes 
\begin{equation}
\label{pi1_harm}
    \pi(x) = \frac{1}{D} \left[ v^2 - \left( D \px +\mu k x\right)^2\right] P(x)    
    - \px \chi(x) \ ,
\end{equation}
where 
\begin{equation}
    \chi(x) = \frac{1}{2v} 
    \left[ \log \frac{(v+\mu k x +D\px)P(x)}{(v-\mu k x-D\px)P(x)} \right] 
    \left[v^2 - \left( D\px+\mu k x\right)^2 \right] P(x) \ .
\end{equation}
We first note that, by integrating over space (\ref{pi1_harm}), we obtain the known expression 
of the global entropy production rate of active particles in harmonic potential 
\cite{paoluzzi2024entropy,Fry2022,GMP2021}.
Indeed, we have
\begin{align}
\Pi &= \int_{-\infty}^{+\infty} dx  \ \pi(x) = \frac{1}{D}\int_{-\infty}^{+\infty} dx \ \left[ v^2-(D\px+\mu k x)^2 \right] P(x) \ \nn \\
&= \frac{1}{D} \int_{-\infty}^{+\infty} dx \int_{-\ell}^{+\ell} dy \ P_A(y) (v^2 - \mu^2 k^2 y^2) P_B(x-y) \nn \\
&= \frac{1}{D} \int_{-\ell}^{+\ell} dy \ P_A(y) (v^2 - \mu^2 k^2 y^2) \nn \\
&= \frac{C}{D} \int_{-\ell}^{+\ell} dy \ (v^2-\mu^2 k^2 y^2)^{\frac{\alpha}{2\mu k}}
= \frac{v^2}{D} \frac{\alpha}{\alpha +\mu k} \ , 
\end{align}
where 
$\ell=v/\mu k$.
In  deriving the above result we
have considered vanishing distributions at boundaries and used
(\ref{P_harm}), (\ref{P_A}), (\ref{P_AA}),  the normalization condition for $P_B$, 
the explicit integral  \cite{gradshteyn2014table}
\begin{equation}
    \int_0^1 dx \ (1-x^2)^\nu = \frac{\sqrt{\pi}}{2} \frac{\Gamma(\nu+1)}{\Gamma(\nu+\frac32)} \ , 
    \qquad \nu>-1, 
\end{equation}
and the property of the Gamma function $\Gamma(z+1)=z \Gamma(z)$.

\noindent
Now, we give a more explicit expression of the local EPR.
By using the convolution expression (\ref{P_harm}), together with ({\ref{P_B}}) and (\ref{P_A}), 
we can write the local EPR (\ref{pi1_harm}) as 
\begin{equation} \label{eprharmexpl}
    \pi(x) = \frac{1}{D} \left[ 
    M_1(x) +\frac{\mu k}{2v} \left(
    M_2(x) \log \frac{Y_+(x)}{Y_-(x)} +
    M_1(x)
    \left(
    \frac{H_+(x)}{Y_+(x)}-\frac{H_-(x)}{Y_-(x)}
    \right)
    \right)
    \right]
\end{equation}
where the different space-dependent quantities are given by
\begin{align}
    M_1(x) &= v^2 I_0(x)- \mu^2 k^2 I_2(x)\ , \\
    M_2(x) &= x M_1(x) - v^2 I_1(x) + \mu^2 k^2 I_3(x)\ , \\
    Y_\pm(x) &= v I_0(x) \pm \mu k I_1(x)\ , \\
    H_\pm(x) &= x Y_\pm(x)  - v I_1(x)\mp \mu k I_2(x)\ ,
\end{align}
where we have defined
\begin{equation}
I_n(x) = I[y^n](x)
\end{equation}
and $I$ is the linear integral operator
\begin{equation}
\label{Idef}
    I[f](x) := \int_{-\ell}^{+\ell} dy \ P_A(y) P_B(x-y) f(y) \ , \qquad \ell=v/\mu k ,
\end{equation}
which, for $f=1$, reduces to the PDF (\ref{P_harm}), i.e., $I[1](x)=I_0(x)=P(x)$.
\\

We can also give an alternative expression for the local EPR, by exploiting the direct form (\ref{pid}).
We first note that the quantities defined in (\ref{Qde}-\ref{JLde}), in the harmonic case $f=-kx$, can be
expressed in terms of the quantities defined above
\begin{align}
    P(x) &= I_0(x) , \\ 
    Q(x) &= \frac{\mu k}{v} I_1(x) , \\
    \Delta(x) &= \frac{M_1(x)}{v} , \\ 
    R(x) &= \frac{Y_+(x)}{2v} , \\ 
    L(x) &= \frac{Y_-(x)}{2v} , \\ 
    J_R(x) &= \frac{M_1(x)}{2v} , \\ 
    J_L(x) &= -\frac{M_1(x)}{2v} .
\end{align}
Substituting in (\ref{pid}) we obtain the alternative expression of $\pi$
\begin{equation}
\label{pihp}
    \pi (x) = \frac{M_1^2(x)}{2vD} \left( \frac{1}{Y_+(x)} + \frac{1}{Y_-(x)} \right) + 
    \frac{\alpha \mu k I_1(x)}{2v} \log \frac{Y_+(x)}{Y_-(x)} .
\end{equation}
The three terms correspond to contributions $\pi_R(x)$, $\pi_L(x)$ and $\pi_{RL}(x)$, see (\ref{pisum}). 
Of course the expressions for $\pi(x)$ in~\eqref{pihp} and in~\eqref{eprharmexpl} are equivalent.

We now compare the theoretical predictions with numerical simulations (in one spatial dimension) of thermal non-interacting RT particles in a harmonic trap. We start with numerical data to highlight the difference in computing $\pi(x)$ 
from Eq.~\eqref{pi_1} (direct expression of $\pi(x)$) and Eq.~\eqref{pi_2} (exploiting $\pi(x)=\phi(x)$ in the steady state). 
In Fig. \ref{fig:rt_harm1}a we adopt (\ref{pi_1}) that we rewrite as
\begin{align} 
    \pi(x) &= \pi_1(x) + \pi_2(x) \\ \nonumber
\pi_1(x) &\equiv \frac{\Delta^2(x)}{4 D}\left(\frac{1}{R(x)} + \frac{1}{L(x)}\right) \\ \nonumber
\pi_2(x) &\equiv \frac{\alpha Q(x)}{2} \log \frac{R(x)}{L(x)} \; .
\end{align}
In Fig. \ref{fig:rt_harm1}b, we compute $\pi(x)$ through (\ref{pi_2}). We obtain that $\chi^\prime \equiv \partial_x \chi(x)$ plays a fundamental role since it is responsible for the peaks in $\pi$ where particles remain stuck. Finally, we compare $P(x)$ from numerics with the numerical integration of the analytical formula and the corresponding $\pi(x)$, as it is shown in \ref{fig:rt_harm2}a and \ref{fig:rt_harm2}b, respectively. 
The density profile $P(x)$ displays a double-peaked structure for $\alpha < \alpha_c$ and a single peak for $\alpha > \alpha_c$, with $\alpha_c=2 \mu k$. This happens because, at small $\alpha$ particles tend to accumulate at distances where external force balances the self-propulsion, while when
$\alpha$ increases, active motion becomes more Brownian-like, and thus particles tend to accumulate at the bottom of the trap. Conversely, the entropy production profile $\pi(x)$ shows a crossover (when increasing $\alpha$) from a triple-peaked structure to a single peak. Here therefore the “inverse monotonicity” relation between density and EPR, observed in the absence of force in Section~\ref{Sec_v}, is clearly violated.  We interpret this  scenario as the result of the complex interplay between the external force and self-propulsion. 

\subsection{Piecewise linear potential}
\label{LinPot}
\begin{figure*}[!t]
\centering\includegraphics[width=1.\textwidth]{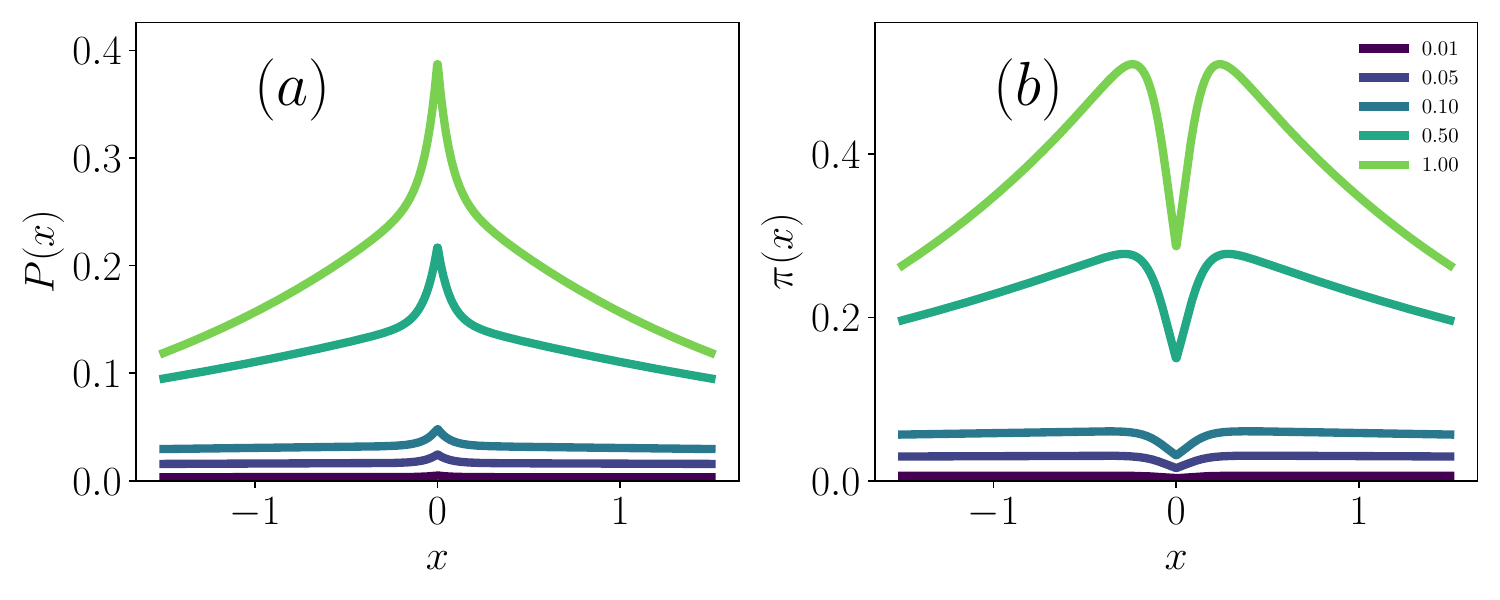}
\caption{Run-and-tumble particle in a linear potential. (a) Stationary distribution $P(x)$ obtained from (\ref{Pst_lin}) ($\alpha \in [0.01,1.0]$, increasing values of $\alpha$ from violet to yellow, as reported in legend of panel (b)), $\mu=v=1$, $f_0=0.5$, and $D=0.1$). (b) The corresponding local EPR $\pi(x)$ from (\ref{epr_lin}).}
\label{fig:rt_lin}      
\end{figure*}

The second case we discuss, which allows an analytical solution, is the piecewise linear potential 
\begin{align}
V(x) &= f_0 |x| \ , \\
f(x)& =\begin{cases}
    - f_0, & \text{if $x>0$} ,\\
    f_0, & \text{if $x<0$}  ,
  \end{cases}
\end{align}
where $f_0>0$ is the force strength.
This case was recently studied in \cite{singh2025}, where an analysis of steady state and relaxation dynamics is reported, along with the expression of the global entropy rate. We discuss here the local EPR in the steady state.
\\
\paragraph*{Stationary solution.}
The steady state equation (\ref{eq_st}) reads
\begin{equation}
\label{SE_lin}
\px \left[ \left(v^2-(D \px +{\text{sgn}}(x) \mu f_0 )^2\right) P(x)\right] + 
\alpha (D \px +{\text{sgn}}(x) \mu f_0 ) P(x) = 0 .
\end{equation}
The solution has the following form
\begin{equation}
\label{P_lin}
    P(x) = A_1 e^{-k_1 |x|} + A_2 e^{-k_2 |x|},
\end{equation}
where the different constants are given in the following (see Appendix \ref{App_Lin} for details).
The quantities $k_1$ and $k_2$ are the absolute values of the two negative solutions of a cubic equation, whose  solutions 
are given by
\begin{equation}
z_n = -\frac23 \frac{\mu f_0}{D} + 2 \sqrt{\frac{p}{3}} \cos{\left[
\frac13 \cos^{-1}\left(
-\frac{3q}{2p}\sqrt{\frac{3}{p}}
\right)
 + \frac23 \pi n
\right]} ,
\qquad {(n=0,1,2)}
\end{equation}
where 
\begin{align}
p &= \frac13 \left( \frac{\mu f_0}{D} \right)^2 + \left( \frac{v}{D}\right)^2 + \frac{\alpha}{D} , \\
q &= \frac{2\mu f_0}{3D} \left[ \left( \frac{v}{D}\right)^2 - \frac{\alpha}{2D} - 
\left( \frac{\mu f_0}{3D} \right)^2 \right] .
\end{align}
The coefficients $A_1$ and $A_2$ are given by
\begin{align}
\label{A1}
    \frac{1}{A_1} &= 2 \left( \frac{1}{k_1} -\frac{1}{k_2} \frac{\mu f_0 - D k_1}{\mu f_0 - D k_2}\right) , \\
    \frac{1}{A_2} &= 2 \left( \frac{1}{k_2} -\frac{1}{k_1} \frac{\mu f_0 - D k_2}{\mu f_0 - D k_1}\right) .
\label{A2}
\end{align}
The PDF (\ref{P_lin}) is shown in Fig. (\ref{fig:rt_lin})a for $\mu=v=1$, $f_0=0.5$, $D=0.1$, and varying $\alpha$.

\begin{figure*}[!t]
\centering\includegraphics[width=1.\textwidth]{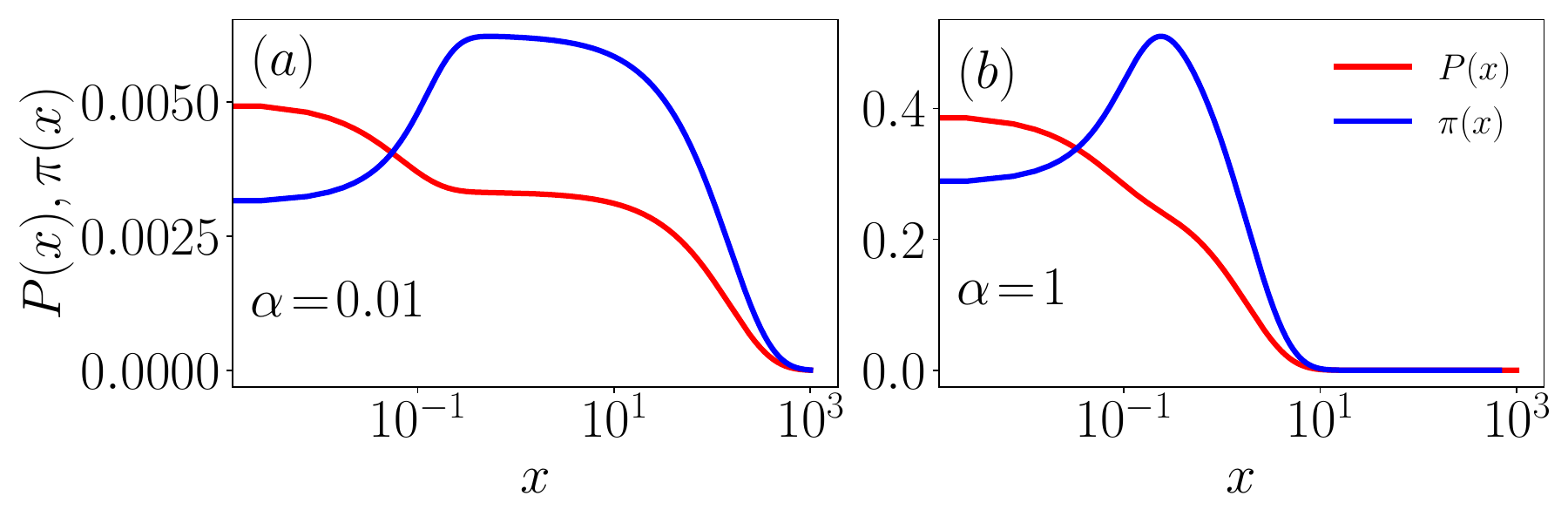}
\caption{Run-and-tumble particle in a linear potential. Two-step decay of $P(x)$ and the corresponding local EPR profile $\pi(x)$ (panel (a) $\alpha=0.01$, panel (b) $\alpha=1$). The other parameters are $D=0.1$, $f_0=0.5$, $\mu=v=1$.}
\label{fig:rt_lin2}      
\end{figure*}

\paragraph*{Entropy production rate.}
We can express the local EPR using one of the expressions previously reported. 
First we note that, for symmetry reasons, $\pi(x)=\pi(-x)$.
By using (\ref{pi_1})  we can write the local ERP as
\begin{equation} \label{epr_lin}
  \pi(x) = \frac{\Delta^2(x)}{D} \frac{P(x)}{P^2(x)-Q^2(x)} + \frac{\alpha Q(x)}{2} \log \frac{P(x)+Q(x)}{P(x)-Q(x)} ,
\end{equation}
where, for $x>0$, we have -- see (\ref{Qde}-\ref{Delde}),
\begin{align}
\label{Pst_lin}
    P(x) &= A_1 e^{-k_1 x} + A_2 e^{-k_2 x} , \\
\label{Qst_lin}
    Q(x) &= \frac{A_1}{v} (\mu f_0 -D k_1) e^{-k_1 x} + 
    \frac{A_2}{v} (\mu f_0 -D k_2) e^{-k_2 x} , \\
    \Delta(x) &= \frac{A_1}{v} \left( v^2 -(\mu f_0 -D k_1)^2 \right) e^{-k_1 x} + 
    \frac{A_2}{v} \left( v^2 -(\mu f_0 -D k_2)^2 \right) e^{-k_2 x} .
\end{align}
We note that the global EPR can be obtained by integrating over the space variable the local quantity.
For this purpose it is convenient to express the local EPR through (\ref{pi_2}), considering vanishing asymptotic distributions.
By using (\ref{Delde}), (\ref{Pst_lin}), (\ref{Qst_lin}) and the property (\ref{A12norm}), we finally obtain 
the expression
\begin{align}
    \Pi &= \int_{-\infty}^{\infty} dx\ \pi(x) =\frac{v}{D} \int_{-\infty}^{\infty} dx\ \Delta(x) = 
    \frac{2v}{D}\int_{0}^{\infty} dx\ \left( v P(x) - \mu f_0 Q(x) \right) \nonumber \\
    &= \frac{v^2-\mu^2 f_0^2}{D} + 2\mu f_0 (A_1+A_2) , 
    \end{align}
in agreement with \cite{singh2025}. In Fig. (\ref{fig:rt_lin})b, we report the local EPR as $\alpha$ changes from $0.01$ to $1$.

In order to highlight the structure of $\pi(x)$, we plot it - together with $P(x)$ - in semilog scale, in Fig.~\ref{fig:rt_lin2}. Again the ``inverse monotonicity'' observed in the cases without external force is violated: the density profile $P(x)$  has always a single peak, as also dictated by the analytical solution, with two different decays at short and long distances from the center; however the EPR $\pi(x)$ displays a complex shape with two symmetric peaks, which are located in a range of distances (from the center)  overlapping with the crossover region between the two slopes of the density decay, a region which is larger for smaller $\alpha$. The interplay between the external force that pushes the particle towards the center and the self-propulsion that creates two populations of particles, some
aligned with the local force and some swimming against it, produces a complex and unexpected dissipation pattern.

\section{Conclusions}
\label{Sec_Conc}
Quantifying deviations from thermodynamic equilibrium on different length scales is of great importance for shedding light into biological processes that usually show a hierarchy of time and length scales. In this work, we have presented a detailed study of the local EPR of run-and-tumble particles. We have shown that it is possible to obtain a closed expression for the local EPR that depends only on the knowledge of the stationary distribution. We get an expression for the local EPR in a general set-up where the 
swimmer parameters can be space-varying functions and external force fields are present. 
As a first result, we showed that, although stationary EPR can be computed using the balance with the entropy flow, once we consider the local quantity, extra care of the gradient terms appearing in the entropy flow is required. 

We have applied our general formalism to analyze in detail two relevant cases which are interesting for their biological or physical relevance.
The first one is that of space-dependent velocity, suitable for describing photokinetic {\itshape E. coli} bacteria,
whose speed can be controlled by the local intensity of projected light patterns into the sample \cite{Elife2018}.
We analyzed two different shapes of the velocity field, the piecewise constant case, where an analytical solution for the stationary distribution in the presence of thermal noise is available, and the case of sinusoidal velocity, for which it is possible to write a perturbation series with thermal diffusion as a small parameter. Numerical simulations corroborate exact and perturbative analytical expressions obtained in the two analyzed cases.
The second situation analyzed is one in which a space-varying external force field is applied.
We focused on the two paradigmatic cases of harmonic and piecewise linear potentials, which allow for exact analytical treatment. 

The analysis of local EPR profiles by varying swimmer parameters, with the presence of crossovers and peaks,
provides interesting  information to characterize the spatial localization of out-of-equilibrium processes in active systems. 

When there is no external force, self-propulsion is the unique source  of energy injection and we observe an ``inverse monotonicity'' relation between $\pi(x)$ and $P(x)$, i.e. the local EPR is larger where the local velocity is larger, corresponding to smaller values of $P(x)$. This is analytically confirmed in the limit of small values of $D$, where one has $\pi(x) \propto v(x)$ and $P(x) \propto 1/v(x)$. On the contrary, the introduction of an external force, leads to a more complex situation that is not easily rationalized even at a qualitative level. Already the density profile exhibits complex behaviors: for instance in the harmonic case $P(x)$ can display a crossover - when reducing the tumbling rate $\alpha$ - from single-peaked to double-peaked; in the presence of a piecewise-linear potential $P(x)$ has a single central peak, with a double-sloped decay. We know that behind this density profiles there are even more complex particle currents $J_L$ and $J_R$, which are combined, in a rich interplay through Eq.~\eqref{therm}, with the local force and the local velocity to give the energy dissipation contribution to EPR; in addition the local EPR is also affected by the local flux $\partial_x J_s(x)$ which is a function of the currents and the densities. In conclusion the local EPR is in general affected by the combined effect of external forces and  self-propulsion, effects  which can constructively or destructively interfere leading to a landscape of maxima or minima. For this reason this quantity is fascinating and further work is needed. 

The local EPR is in our opinion a quantity that is still in its infancy as an interesting tool for theory and experiments. Our opinion is that measuring the distribution in space of the dissipation of a system is more informative than a global quantity, when the system’s volume is not homogeneous, for instance in a living cell (see for instance \cite{di2024variance})
in active systems undergoing motility-induced phase separation or in the presence of rectifying devices \cite{PhysRevLett.129.220601}.


In some situations of practical interest, as in the case of photokinetic bacteria, the expressions we obtained can be employed directly in experiments for visualizing the local EPR field. This is because in photokinetics one controls the velocity field and can compute numerically the stationary distributions. As a future direction, it would be interesting to extend this computation in the case of a density-dependent velocity to include excluded volume effects and motility-induced phase separation. 
We also stress that the computation we presented here can be extended in $d$ spatial dimensions.

\begin{acknowledgments}
L.A. acknowledges funding from the Italian Ministero dell’Università e della Ricerca under the programme
PRIN 2020, number 2020PFCXPE.
M.P. and A.P. acknowledge funding from the Italian Ministero dell’Università e della Ricerca under the programme PRIN 2022 ("re-ranking of the final lists"), number 2022KWTEB7, cup B53C24006470006.
\end{acknowledgments}


\appendix

\section{Numerical simulations}
\label{App_Sim}
Numerical simulations are performed considering a gas of independent run-and-tumble particles in 1D 
obeying the overdamped Langevin equation of motion
\begin{equation}
\label{Lang_eq}
    \dot{x} = v(x) e +\mu f(x) +\eta ,
\end{equation}
where $\dot{x}$ indicates the time derivative, the direction of motion $e= \pm 1$ is a dichotomous random variable updated at rate $\alpha$ (tumbling rate), $v(x)$ is the particle speed
(which can be a space-dependent function), $f(x)$ is a generic force field, $\mu$ is the particle mobility, and $\eta$ is the
thermal noise with $\langle \eta(t)\rangle=0$ and $\langle \eta(t)\eta(\tau)\rangle = 2D\delta(t-\tau)$, with $D$ the thermal diffusion constant.
The equation (\ref{Lang_eq}) is numerically integrated by using the Euler scheme with time step $\Delta t=10^{-2}$, considering $N=10^3$ independent runners for a number of $N_T=10^{6}$ time steps. To compute the corresponding probability density function $P(x)$, we collect stationary trajectories considering the second half time steps for each runner. The local entropy production rate $\pi(x)$ is thus computed using (\ref{pi_1}) and (\ref{pi_2}), where we compute the various stationary fields, i.e., $Q(x)$, $\Delta(x)$, and $J_{R,L}(x)$, through (16)-(21) using their relations with $P(x)$, $f(x)$, and $v(x)$.


\section{Stationary PDF of RT particles in a thermal bath with a piecewise constant speed}
\label{App_Speed}
The PDFs in the two zones, $0<x<\lambda$, where $v(x)=v_1$, and  $\lambda<x<L$, where $v(x)=v_2$,   
have the form 
\begin{equation}
\label{P_i}
    P_i(x) = A_i e^{k_i x} + B_i e^{-k_i x} + C_i , \qquad (i=1,2)
\end{equation}
where
\begin{equation}
    D^2 k_i^2 = v_i^2 +\alpha D .
\end{equation}
The six unknown constants ($A_i,B_i,C_i$), with $i=1,2$, are determined by imposing the following conditions on $P_i(x)$
\begin{align}
& P_1(x) = P_1(\lambda -x) , \qquad &\text{for }  x\in(0,\lambda) \qquad &\text{(symmetry in zone 1)} \label{AC1}\\
& P_2(x) = P_2(L+\lambda -x) , \qquad &\text{for } x\in(\lambda,L) \qquad &\text{(symmetry in zone 2)} \label{AC2}\\
& P_1(0) =P_2(L) , & & \text{(continuity of $P$)}   \label{AC3}\\
& \left. \frac{\px P_1(x)}{v_1}\right|_{x=0} = \left. \frac{\px P_2(x)}{v_2}\right|_{x=L} , & &  \text{(continuity of $Q$)}  \label{AC4}\\
& \left[ v_1 P_1(x) - \frac{D^2}{v_1} \px^2 P_1(x)\right]_{x=0} = 
 \left[ v_2 P_2(x) - \frac{D^2}{v_2} \px^2 P_2(x)\right]_{x=L}   & &  \text{(continuity of $\Delta$)}  \label{AC5}\\
& \int_0^\lambda dx\ P_1(x) + \int_\lambda^L dx\ P_2(x) =1, & &  \text{(normalization)}  \label{AC6}
\end{align}
Continuity conditions in $x=\lambda$ are not independent from that on $x=0$, due to the symmetry conditions 
(\ref{AC1}) and (\ref{AC2}). See (\ref{QRL}) and (\ref{DeltaDef}) for the definition of $Q$ and $\Delta$ quantities.
By substituting (\ref{P_i}) in the above equations we get a set of linear equations for the unknown constants.
After some algebra we finally obtain  the expressions
\begin{align}
    A_1 &= \frac{1}{Y_1 - Y_2 - Y_3  Y_4} , \\
    A_2 &= A_1 \frac{v_2 k_1}{v_1 k_2} \frac{1-e^{k_1\lambda}}{e^{k_2L}-e^{k_2\lambda}} , \\
    B_1 &= A_1 e^{k_1\lambda} , \\
    B_2 &= A_2 e^{k_2(L+\lambda)} , \\
    C_1 &= \frac{1-A_1 (Y_3(L-\lambda)+Y_1)}{L} , \\
    C_2 &= \frac{1+A_1 (Y_3 \lambda -Y_1)}{L} ,   
\end{align}
where
\begin{align}
    Y_1 &= \frac{2}{k_1} (e^{k_1\lambda}-1) \left( 1-\frac{v_2 k_1^2}{v_1 k_2^2}\right) , \\
    Y_2 &= \frac{\alpha L D}{v_1 v_2} (1+e^{k_1 \lambda}) , \\
    Y_3 &= 1+e^{k_1 \lambda} - \frac{v_2 k_1}{v_1 k_2} \frac{(1-e^{k_1\lambda})(e^{k_2L}+e^{k_2\lambda})}{e^{k_2L}-e^{k_2\lambda}}  , \\
    Y_4 &= \lambda + \frac{L}{v_2-v_1} \left( \frac{\alpha D}{v_2}+v_1\right) .
\end{align}
In the limit of equal speeds in the two zones, $v_1=v_2$, we have that $Y_4 \to \infty$, resulting in vanishing 
$A_i$ and $B_i$, and $C_i=1/L$ for $i=1,2$. Therefore we obtain the trivial constant PDF $P(x)=1/L$.

\section{Stationary PDF of RT particles in a thermal bath in harmonic potential}
\label{App_Harm}
We demonstrate that the PDF (\ref{P_harm}), that we report again here for convenience,
\begin{equation}
\label{PhA}
    P(x) = \int_{-\ell}^{+\ell} dy \ P_A(y) \ P_B(x-y) \ , \qquad \ell=v/\mu k \ ,
\end{equation}
with $P_B$ given in (\ref{P_B}) and solution of (\ref{eq_PB}),
$P_A$ given in (\ref{P_A}) and solution of (\ref{eq_PA}), is solution of the stationary equation 
(\ref{eq_st}) with $f(x)=-kx$.
We first note that, using (\ref{eq_PB})
\begin{align}
    [D\px-\mu f(x)] P_B(x-y) &= \mu [f(x-y)-f(x)] P_B(x-y) 
    \nonumber \\
    &= \mu k y P_B(x-y) \ ,
\end{align}
and 
\begin{equation}
    [D\px-\mu f(x)]^2 P_B(x-y) = (\mu k y)^2 P_B(x-y) \ .
\end{equation}
Then, inserting (\ref{PhA}) in (\ref{eq_st}), we obtain 
\begin{align}
 &   \int_{-\ell}^{+\ell} dy \ P_A(y) \left[
    \px \left(
    (v^2-\mu^2 k^2 y^2) P_B(x-y) 
    \right)
    +\alpha \mu k y P_B(x-y) 
    \right] \nonumber  \\
  = &  \int_{-\ell}^{+\ell} dy \  P_A(y)  (v^2-\mu^2 k^2 y^2)\px  P_B(x-y) 
    +\alpha \mu k \int_{-\ell}^{+\ell} dy \ y P_A(y) P_B(x-y) \nonumber \\
  = & - \int_{-\ell}^{+\ell} dy \ P_A(y)  (v^2-\mu^2 k^2 y^2) \py P_B(x-y) 
    +\alpha \mu k \int_{-\ell}^{+\ell} dy \ y P_A(y) P_B(x-y) \nonumber \\
   = & P_A(y) (v^2-\mu^2 k^2 y^2)P_B(x-y)\Big{|}_{y=-\ell}^{y=\ell}
   +\int_{-\ell}^{+\ell} dy \ P_B(x-y) \py [(v^2-\mu^2 k^2 y^2) P_A(y)] \nonumber \\
   & + \alpha \mu k \int_{-\ell}^{+\ell} dy \ y P_A(y) P_B(x-y) \nonumber \\
  = & - \alpha \mu k \int_{-\ell}^{+\ell} dy \ y P_A(y) P_B(x-y)  
    +\alpha \mu k \int_{-\ell}^{+\ell} dy \ y P_A(y) P_B(x-y)  = 0 \ ,
\end{align}
where boundary terms are zero and we have used (\ref{eq_PA}).
Therefore, the PDF (\ref{PhA}) is the stationary solution of  the run-and-tumble motion
with thermal noise in the presence of harmonic potential.

\section{Stationary PDF of RT particles in a thermal bath in a piecewise linear potential}
\label{App_Lin}
The equation of the stationary PDF $P(x)$ of a thermal RT particle in a piecewise lienar potential 
$V(x)=f_0 |x|$ (with $f_0>0$) is the third order differential equation (\ref{SE_lin}).
For symmetry reasons, we have $P(-x)=P(x)$ and we consider here the case $x>0$. 
We can write the differential equation in the form
\begin{equation}
    \label{3DE}
    P''' + aP'' +b P' -c =0 ,
\end{equation}
where $'$ indicates the space derivative and 
\begin{align}
a &= \frac{2\mu f_0}{D} > 0 ,\\
b &= \left( \frac{\mu f_0}{D}\right)^2 - \left(\frac{v}{D} \right)^2 - \frac{\alpha}{D}  , \\
c &= \frac{\alpha \mu f_0}{D^2} > 0 .
\end{align}
The corresponding characteristic cubic equation is
\cite{Nickalls_1993,Nickalls_2006,Zucker2008,singh2025}
\begin{equation}
\label{ceqz}
    h(z)= z^3 + a z^2 + bz -c =0 ,
\end{equation}
which, by changing variable $y = z+a/3$, leads to the reduced equation
\begin{equation}
\label{ceqy}
g(y)=    y^3 -p y + q =0 ,
\end{equation}
where $p$ and $q$ are
\begin{align}
    p &=\frac{a^2}{3} - b =\frac13 \left( \frac{\mu f_0}{D}\right)^2 + \left(\frac{v}{D} \right)^2 + \frac{\alpha}{D} >0 , \\
    q &=\frac{2a^3}{27} - \frac{ab}{3} -c = \frac{2\mu f_0}{3D} \left[  \left(\frac{v}{D} \right)^2 - \frac{\alpha}{D} -\left( \frac{\mu f_0}{3D}\right)^2 
\right] .
\end{align}
The discriminant $\delta=4 p^3 - 27 q^2$ detemines the nature of the roots of the cubic equation.
In our case we have
\begin{equation*}
\delta = 4 \left( \frac{v}{D}\right)^2 \left[ \left( \frac{\mu f_0}{D}\right)^2 - \left( \frac{v}{D}\right)^2\right]^2 +
\frac{\alpha}{D} \left[ 
20 \left( \frac{v \mu f_0}{D^2}\right)^2 + 12 \left( \frac{v}{D} \right)^4 + \frac{\alpha}{D}\left( \frac{\mu f_0}{D}\right)^2
+ 12 \frac{\alpha}{D} \left( \frac{v}{D}\right)^2 + 4 \left( \frac{\alpha}{D}\right)^2 
\right] >0 ,
\end{equation*}
and then equation (\ref{ceqy}) has three distinct real solutions
(the stationary points $y_{\pm} = \pm \sqrt{p/3}$ of the cubic form $g(y)$ are such that $g(y_+)g(y_-)=- \delta/27<0$).
These can be found by writing $$y=u \cos \varphi$$ and setting $u=2\sqrt{p/3}$. By using the trigonometric identity 
$\cos{3\varphi} = 4 \cos^3{\varphi} -3 \cos{\varphi}$, equation (\ref{ceqy}) takes the form 
\begin{equation}
2 \left( \frac{p}{3}\right)^{3/2} \cos{3\varphi} + q = 0 ,
\end{equation}
with solutions
\begin{equation}
    \varphi_n = \frac13 \cos^{-1} \left( - \frac{3q}{2p}\sqrt{\frac{3}{p}}\right) + \frac{2\pi n}{3} , \qquad (n=0,1,2) .
\end{equation}
The three solutions of (\ref{ceqy}) can then be written as
\begin{equation}
   y_n = 2 \sqrt{\frac{p}{3}} \cos \left[ 
   \frac13 \cos^{-1} \left( - \frac{3q}{2p}\sqrt{\frac{3}{p}}\right) + \frac{2\pi n}{3} 
   \right] , \qquad (n=0,1,2) 
\end{equation}
Therefore, also the equation (\ref{ceqz}) has three distinct solutions, two negatives and one positive, 
as follows from the fact that $h(z=0)=-c<0$, $\lim_{z\to \pm \infty} g(z) = \pm \infty$
and the flex point at which the function $h(z)$ changes curvature is at $z_s=-a/3<0$.
The solutions are 
\begin{equation}
\label{zn}
    z_n = -\frac{2\mu f_0}{3D} + y_n , \qquad (n=0,1,2) .
\end{equation}
The solutions of the differential equation (\ref{3DE}) can be written as a linear combination 
\begin{equation}
    P(x) = A_1 e^{-k_1 x} + A_2 e^{-k_2 x} , 
\end{equation}
where $k_1$ and $k_2$ are the absolute values of the two negative solutions in (\ref{zn}) (we require that $P(x)$ does not diverge
at large $x$). The coefficients $A_i$ ($i=1,2$) can be found by imposing normalization of $P$, $\int_0^\infty P(x) dx =1/2$,
and vanishing of $Q(x)=R(x)-L(x) = (D/v)\px P(x) + (\mu f_0/v)P(x)$ at $x=0$ (symmetry). We obtain
\begin{align}
\label{A12norm}
     \frac{A_1}{k_1} + \frac{A_2}{k_2} &= \frac{1}{2} , \\
    A_1(\mu f_0 - D k_1) &= -A_2 (\mu f_0 -D k_2) ,
\end{align}
leading to the expressions (\ref{A1}) and (\ref{A2}).
\section{Notes on different definitions of entropy}
\label{App_Entropies}
Here we explore different definitions of entropy, which generalize the Gibbs entropy used before \cite{Amigo2018}.

\subsection{R\'enyi Entropy}
Given a PDF $P(x,t)$ the R\'enyi entropy $S^R_q$ is defined as follows 
\cite{Renyi1960,Ozawa_2024}
\begin{equation}
    \label{RE}
    S^R_q(t) = \frac{1}{1-q} \log \int dx\ P^q(x,t)  ,
\end{equation}
where $q$ is an index with $q\in(0,+\infty)$.
In the limit $q\to 1$ the R\'enyi entropy tends to the Gibbs entropy
\begin{equation}
    \lim_{q\to 1} S^R_q(t) = - \int dx \ P(x,t) \log P(x,t) .
\end{equation}
For the R\'enyi entropy we cannot define a local entropy 
, but we will be able to introduce a local EPR (see below).\\

In the case of RT particles we can use the above definitions by substituting
$\int dx \to \sum \int dx$, where the sum runs over the internal states ($R$ and $L$
in our 1d case).
We obtain the following definition of R\'enyi entropy for RT particles
\begin{equation}
    S^R_q(t) = \frac{1}{1-q} \log \int dx \ \left( R^q(x,t) + L^q(x,t) \right) .
\end{equation}
The entropy production rate can be calculated using the same procedure followed in the previous sections.
We have that 
\begin{equation}
    \pt S^R_q(t) = \frac{q}{(1-q)N_q(t)} \int dx \ \left( 
    R^{q-1}(x,t)\pt R(x,t) +   L^{q-1}(x,t)\pt L(x,t) 
    \right) ,
\end{equation} 
where 
\begin{equation}
\label{Nq}
    N_q(t) = \int dx \ \left( R^q(x,t) + L^q(x,t) \right) = e^{(1-q) S^R_q(t) }.
\end{equation}
We note that $N_q$ depends only on the time variable (in the stationary regime is constant) and 
in the limit $q\to 1$ we have $N_q \to 1$, resulting in the equivalence between the R\'enyi and Gibbs 
entropy rate in this limit.
We can define a local entropy rate as follows
\begin{align}
    \pt s^R_q(x,t) &= \frac{q}{(1-q)N_q(t)}  \left( 
    R^{q-1}(x,t)\pt R(x,t) +   L^{q-1}(x,t)\pt L(x,t) 
    \right)  + \frac{\pt P(x,t)}{(q-1)N_q(t)} \nonumber \\
    & = 
 \frac{1}{(q-1)N_q(t)}  \left[ \left(1-qR^{q-1}(x,t)\right) \pt R(x,t)  + 
  \left( 1-qL^{q-1}(x,t)\right)\pt L(x,t) 
    \right]   , 
\label{LRER}
\end{align}
having used the freedom to add a term whose space intergral is zero (due to the normalization of $P=R+L$).
We chose this expression because it guarantees that, in the limit $q\to 1$, we recover the
correct results of the Gibbs case for local entropy rates (see below).
After some algebra, we finally arrive at the expression
\begin{equation}
    \pt s^R_q(x,t) =  \pi^R_q(x,t) - \phi^R_q(x,t) ,
\end{equation}
where the first term is the local R\'enyi EPR (non negative quantity) defined by
\begin{align}
\label{piR}
 \pi^R_q(x,t) &= \frac{q}{DN_q(t)} \left( J_R^2(x,t) R^{q-2}(x,t) + J_L^2(x,t) L^{q-2}(x,t)\right) \\
    &+ \frac{\alpha}{2N_q(t)} \frac{q}{q-1} (R(x,t)-L(x,t)) \left( R^{q-1}(x,t) - L^{q-1}(x,t)\right) , 
\end{align}
and the second term is the local R\'enyi entropy flux
\begin{align}
\label{phiR}
 \phi^R_q(x,t) &= \frac{qv}{DN_q(t)} \left( J_R(x,t) R^{q-1}(x,t) - J_L(x,t) L^{q-1}(x,t)\right) \\
     &+ \frac{q\mu f(x)}{DN_q(t)} \left( J_R(x,t) R^{q-1}(x,t) + J_L(x,t) L^{q-1}(x,t)\right) 
     - \frac{1}{N_q(t)}\px \chi(x,t) ,
\end{align}
where  
\begin{equation}
\label{chiT}
    \chi(x,t) = \frac{J_R(x,t) (q R^{q-1}(x,t)-1) +J_L(x,t) (q L^{q-1}(x,t)-1) }{q-1} .
\end{equation}
For $q \to 1$ the above expressions reduce to those found before, (\ref{pid}), (\ref{phid}) and (\ref{chid}),
referring to the Gibbs entropy.\\
The global EPR $\Pi^R_q$ and entropy flux $\Phi^R_q$ are obtained as space integral of the corresponding local quantities.\\
In the stationary regime we have that
\begin{equation}
 \pi^R_q(x) = \phi^R_q(x) , 
\end{equation}
and, then, we can calculate the local EPR 
through (\ref{piR}) or (\ref{phiR}),
once we know the stationary PDF $P(x)$, solution of (\ref{SPDFeq}),
and using  (\ref{Qde}-\ref{JLde}).
\\

We can give an explicit expression of the local R\'enyi EPR 
for the case of harmonic potential 
in terms of the quantities defined at the end of section \ref{HarPot}.
We obtain
\begin{equation}
\label{pihpR}
    \pi^R_q(x) = \frac{q}{(2v)^q N_q} \left[
 \frac{M_1^2(x)}{D} \left( Y_+^{q-2}(x) + Y_-^{q-2}(x) \right) +
 \frac{\alpha \mu k I_1(x)}{q-1} \left( Y_+^{q-1}(x) - Y_-^{q-1}(x)\right)
 \right] ,
\end{equation}
where 
\begin{equation}
    N_q = \int \frac{dx}{(2v)^q} \ \left( Y_+^q(x) + Y_-^q(x) \right) .
\end{equation}
For $q\to 1$ we recover the Gibbs expression (\ref{pihp}).

\subsection{Tsallis Entropy}
Given a PDF $P(x,t)$ the Tsallis entropy $S^T_q$ is defined as follow \cite{Tsallis1988}
\begin{equation}
    \label{TE}
    S^T_q(t) = \frac{1}{q-1} \left( 1 - \int dx\ P^q(x,t) \right) ,
\end{equation}
where $q$ is an index with $q\in(0,+\infty)$.
The Tsallis entropy is known to be a non-additive quantity.
We can define the local ($x$-dependent) entropy as
\begin{equation}
    \label{LTE}
    s^T_q(x,t) = \frac{P(x,t) - P^q(x,t)}{q-1} .
\end{equation}
In the limit $q \to 1$ we recover the Gibbs entropy
\begin{align}
\lim_{q \to 1} s^T_q(x,t) &= - P(x,t) \log P(x,t) , \\
\lim_{q \to 1} S^T_q(t) &= - \int dx \ P(x,t) \log P(x,t) .
\end{align}
It is worth noting that the Tsallis ans R\'enyi entropies are related one each other
through  
\begin{align}
    S^T_q(t) &= \frac{e^{(1-q)S^R_q(t)}-1}{1-q}  , \\
    S^R_q(t) &= \frac{1}{1-q} \log \left( 1 + (1-q) S^T_q(t)\right) .
\end{align}

In the case of RT particles we can define local and global Tsallis entropies as follows
\begin{align}
    s^T_q(x,t) &= \frac{1}{q-1} \left( R(x,t) - R^q(x,t) + L(x,t)-L^q(x,t)  \right) , \\
    S^T_q(t) &= \frac{1}{q-1} \left( 1 - \int dx\ \left( R^q(x,t) + L^q(x,t) \right) \right) .
\end{align}
For the local entropy production rate we have that 
\begin{equation}
    \pt s^T_q(x,t) = \frac{1}{q-1} \left[ 
    (1-qR^{q-1}(x,t))\pt R(x,t) +     (1-qL^{q-1}(x,t))\pt L(x,t) 
    \right] = N_q(t) \pt s^R_q(x,t) , 
\end{equation}
i.e., it can be expressed in terms of the local R\'enyi entropy rate (\ref{LRER}).
We then obtain the following expression
\begin{equation}
    \pt s^T_q(x,t) = \pi^T_q(x,t) - \phi^T_q(x,t) , 
\end{equation}
where the local Tsallis EPR and entropy flux are expressed in terms of (\ref{piR}) and (\ref{phiR})
\begin{align}
\label{piT}
    \pi^T_q(x,t) &= N_q(t) \pi^R_q(x,t) , \\ 
    \phi^T_q(x,t) &= N_q(t) \phi^R_q(x,t) .
\label{phiT}
\end{align}
Therefore, the local Tsallis entropy rates are proportional to the corresponding R\'enyi ones 
with a proportionality (time dependent) factor $N_q(t)$, given in (\ref{Nq}).
The global EPR $\Pi^T_q$ and entropy flux $\Phi^T_q$ are obtained as space integral of the corresponding local quantities.
\\
As before, in the stationary regime we have that
\begin{equation}
 \pi^T_q(x) = \phi^T_q(x) ,
\end{equation}
and we can use one of them to calculate the EPR once we know the exact expression of the PDF $P(x)$.
\\

The explicit expression of the stationary local Tsallis EPR in the case of harmonic potential
is obtained from (\ref{piT}) and (\ref{pihpR}), resulting 
in the same expression of the R\'enyi EPR rescale by the factor $N_q$
\begin{equation}
\label{pihpT}
    \pi^T_q(x) = \frac{q}{(2v)^q} \left[
 \frac{M_1^2(x)}{D} \left( Y_+^{q-2}(x) + Y_-^{q-2}(x) \right) +
 \frac{\alpha \mu k I_1(x)}{q-1} \left( Y_+^{q-1}(x)- Y_-^{q-1}(x)\right)
 \right] .
\end{equation}

%

\end{document}